\documentclass[twocolumn]{aastex63}

\usepackage[super]{nth}
\usepackage{savesym}
\savesymbol{tablenum}
\usepackage[T1]{fontenc}
\usepackage{inputenc}
\usepackage[range-units = brackets,tophrase={-},seperr,repeatunits=false]{siunitx}
\usepackage{multirow}
\restoresymbol{SIX}{tablenum}

\graphicspath{{./}{figures/}}

\submitjournal{ApJ}

\shorttitle{The merging cluster Abell~2108}
\shortauthors{Schellenberger et al.}

\begin{document}

\title{The unusually weak and exceptionally steep radio relic in Abell~2108}
\correspondingauthor{Gerrit Schellenberger}
\email{gerrit.schellenberger@cfa.harvard.edu}

\author[0000-0002-4962-0740]{Gerrit Schellenberger}
\affiliation{Center for Astrophysics $|$ Harvard \& Smithsonian, 60 Garden St., Cambridge, MA 02138, USA}

\author[0000-0002-1634-9886]{Simona Giacintucci}
\affiliation{Naval Research Laboratory, 4555 Overlook Avenue SW, Code 7213, Washington, DC 20375, USA}

\author[0000-0002-3754-2415]{Lorenzo Lovisari}
\affiliation{INAF - Osservatorio di Astrofisica e Scienza dello Spazio di Bologna, via Piero Gobetti 93/3, 40129 Bologna, Italy}
\affiliation{Center for Astrophysics $|$ Harvard \& Smithsonian, 60 Garden St., Cambridge, MA 02138, USA}

\author[0000-0002-5671-6900]{Ewan O'Sullivan}
\affiliation{Center for Astrophysics $|$ Harvard \& Smithsonian, 60 Garden St., Cambridge, MA 02138, USA}

\author{Jan Vrtilek}
\affiliation{Center for Astrophysics $|$ Harvard \& Smithsonian, 60 Garden St., Cambridge, MA 02138, USA}

\author{Laurence P. David}
\affiliation{Center for Astrophysics $|$ Harvard \& Smithsonian, 60 Garden St., Cambridge, MA 02138, USA}

\author{Jean-Baptiste Melin}
\affiliation{IRFU, CEA, Université Paris-Saclay, 91191 Gif-sur-Yvette, France}

\author[0000-0001-5470-305X]{Dharam Vir Lal}
\affiliation{National Centre for Radio Astrophysics - Tata Institute of Fundamental Research, Ganeshkhind P.O., Pune 411007, India}

\author[0000-0003-4117-8617]{Stefano Ettori}
\affiliation{INAF - Osservatorio di Astrofisica e Scienza dello Spazio di Bologna, via Piero Gobetti 93/3, 40129 Bologna, Italy}
\affiliation{INFN, Sezione di Bologna, viale Berti Pichat 6/2, 40127 Bologna, Italy}

\author[0000-0002-3104-6154]{Konstantinos Kolokythas}
\affiliation{Centre for Space Research, North-West University, Potchefstroom 2520, South Africa}

\author[0000-0003-0302-0325]{Mauro Sereno}
\affiliation{INAF - Osservatorio di Astrofisica e Scienza dello Spazio di Bologna, via Piero Gobetti 93/3, 40129 Bologna, Italy}

\author[0000-0002-4864-4046]{Somak Raychaudhury}
\affiliation{Inter-University Centre for Astronomy and Astrophysics, Savitribai Phule Pune University Campus, Ganeshkhind, Pune 411007, India}
\affiliation{School of Physics and Astronomy, University of Birmingham, Birmingham B15~2TT, UK}

\begin{abstract}
Mergers between galaxy clusters often drive shocks into the intra cluster medium (ICM), the effects of which are sometimes visible via temperature and density jumps in the X-ray, and via radio emission from relativistic particles energized by the shock's passage.
Abell~2108 was selected as a likely merger system through comparing the X-ray luminosity to the Planck Sunyaev-Zeldovich signal, where this cluster appeared highly X-ray underluminous. 
Follow up observations confirmed it to be a merging low mass cluster featuring two distinct subclusters, both with a highly disturbed X-ray morphology. 
Giant Metrewave Radio Telescope (GMRT) data in bands 2, 3 \& 4 (covering 120-750\,MHz) show an extended radio feature resembling a radio relic, near the location of a temperature discontinuity in the X-rays. 
We measure a Mach number from the X-ray temperature jump ($\mathcal{M}_\text{X} = \num{1.6(2)}$).
Several characteristics of radio relics (location and morphology of extended radio emission) are found in Abell~2108, making this cluster one of the few low mass mergers  ($M_\text{L-M}=\SI{1.8(4)e14}{M_\odot}$) likely hosting a radio relic. The radio spectrum is exceptionally steep ($\alpha = -2$ at the lowest frequencies), and the radio power is very weak ($P_{\SI{1.4}{GHz}} = \SI{1e22}{W\,Hz^{-1}}$). 
To account for the shock/relic offset, we propose a scenario in which the shock created the relic by re-accelerating a cloud of pre-existing relativistic electrons and then moved away, leaving behind a {\it fading} relic. The electron aging timescale derived from the high-frequency steepening in the relic spectrum is consistent with the shock travel time to the observed X-ray discontinuity. 
However, the lower flux in GMRT band 4 data causing the steepening could be due to instrumental limitations,
and deeper radio data are needed to constrain the spectral slope of the relic at high frequencies. 
A background cluster, 4\,arcmin from the merger, may have contributed to the ROSAT and Planck signals, but SZ modeling shows that the merger system is still X-ray underluminous, supporting the use of this approach to identifying merger-disrupted clusters.

\end{abstract}


\section{Introduction} \label{sec:intro}
Galaxy cluster mergers are the most energetic events in the Universe. They are predicted from hierarchical structure formation theory, and are ideal laboratories to study gravitational mass assembly. However, while rich and massive clusters are very rare compared to low mass systems ($<\SI{2e14}{M_\odot}$), most of the violent mergers which have been studied are in high mass systems.
Merging galaxy groups and low mass clusters provide a much more representative view on the population of mergers, and due to the lower gravitational potential, are more affected by non-gravitational effects of mergers and active galactic nuclei (AGNs, see e.g., \citealp{Forbes2006-ay,Sun2012-ll}).
While galaxy groups have already proven to be of extraordinary importance when understanding feedback processes, scaling relations, and hydrostatic equilibrium violations  (e.g., \citealp{Sun2009-pg,Lovisari2015-xm,Schellenberger2017-mc,Lovisari2021-ix,Eckert2021-mm}), most cosmological applications rely on galaxy clusters and the precise knowledge of observable-mass scaling relations. To be able to include low mass systems in this picture (especially in the light of the eROSITA survey, \citealp{Merloni2012-sc}), substantial understanding of the impact of mergers on the galaxy and gas environment is crucial.

Galaxy cluster mergers cause shocks which can heat the gas to high temperatures, but so far only a few mergers of low mass clusters and groups are known (e.g., \citealp{Machacek2005-el,Machacek2010-qu,Machacek2011-po,Kraft2006-wn,Gastaldello2013-cp,OSullivan2014-ku,OSullivan2019-le}). 
Although expected, shock fronts and/or radio relics (diffuse synchrotron sources that trace the shock waves) are especially rare among these objects (\citealp{Randall2009-dn,Russell2014-zj,Van_Weeren2019-qg}). 
A systematic search in Abell clusters showed that only 2\% have good indications for relics or halos (\citealp{Kempner2001-eb}), and even in radio selected samples only about 20-30\% of the sources turn out to be relics (\citealp{Van_Weeren2009-sn}), while twice the number of relics are expected based on simulations (\citealp{Nuza2017-ir}). 
Particle scattering up- and downstream of the shock by plasma irregularities (Diffusive shock acceleration, DSA) is thought to be the main mechanisms to produce radio relics (\citealp{Brunetti2014-lw,Locatelli2020-cs}). However, observations of Mach numbers of shocks raise doubts that this mechanism really works in practice and reacceleration of seed relativistic electrons in the ICM may be required to explain the radio properties of relic (e.g., \citealp{Vazza2014-ta,Van_Weeren2019-qg,Botteon2020-co}).

Comparing the ROSAT (soft) X-ray flux ($\propto n_\text{e}^2$, where $n_\text{e}$ is the electron density) and the Planck Sunyaev-Zeldovich (SZ) signal tracing the gas pressure ($\propto n_\text{e} T$, where $T$ is the ICM temperature) can help to identify merging clusters, since shock heating of the ICM will produce a greater increase in SZ brightness than in X-rays. 
Our goal is to characterize the X-ray under-luminous cluster population, where we classify a cluster as being X-ray underluminous when its SZ predicted X-ray luminosity exceeds 150\% of measured X-ray luminosity. Exploring these X-ray underluminous objects provides insights in a population of clusters, which likely has a very high major merger rate (e.g., \citealp{Popesso2007-hp}), and offers the possibility to challenge models for the re-distribution of kinetic energy to thermal and non-thermal particles. Hence, we select a sample of clusters with low X-ray luminosity and strong SZ signals. For all of our 15 objects in the sample the derived mass from the SZ signal is at least 50\% larger than the X-ray luminosity mass.
In our pilot study we found Abell~2108 to be highly X-ray underluminous in a comparison of catalog masses of Planck SZ clusters and ROSAT X-ray, and follow-up observations revealed the complicated structure. 

This paper is structured as follows. In Section \ref{ch:observations}, we describe the X-ray, and GMRT radio data sets along with the reduction steps. In section \ref{ch:results}, we lay out the structure of the system, describing the two parts of Abell~2108 and the interesting features within, along with more details on the background cluster that is observed in the field of view. Section \ref{ch:discussion} discusses the results, in particular we compare the shock strength to other known merging clusters, and describe how this object is likely the least massive merger hosting a radio relic known today. Section \ref{ch:summary} summarizes the findings. 

Throughout this paper we assume a flat $\Lambda$CDM cosmology with the following parameters: $\Omega_{\rm m} = \num{0.3}$, $\Omega_\Lambda = \num{0.7}$, $H_0 =  h \cdot \SI{100}{km~s^{-1}~Mpc^{-1}}$ with $h = \num{0.7}$, which gives an angular scale of $\SI{1.71}{kpc}$ per arcsec at redshift 0.092 of Abell~2108. 
Uncertainties are stated at the 68\% confidence level, and cluster masses refer to a radius, within which the mean density of the cluster is 500 times the critical density of the Universe at the cluster redshift.

\section{Observations}
\label{ch:observations}
Our radio and X-ray observations of Abell~2108 (J2000 coordinates R.A.$15^{\rm{h}}40^{\rm{m}}01^{\rm{s}}$ Decl.$+17^{\rm d}52^{\rm m}41^{\rm s}$) are essential to reveal the characteristics of this system, and examine why SZ and X-ray measurements deviate strongly from expectations here. In the following we describe the important data reduction and analysis steps.
 
\subsection{Chandra and XMM-Newton X-ray data}
\begin{deluxetable*}{cccccccc}
	\tablecaption{Summary of the X-ray observations \label{tab:xraydata}}
	\tablehead{
		\colhead{Instrument} & \colhead{Obsid} & \colhead{PI} & \colhead{Date} & \colhead{Clean  time} & \colhead{Resolution} & \colhead{Filter} & \colhead{Readout mode}\\
		\colhead{} & \colhead{} & \colhead{} & \colhead{} & \colhead{(ks)} & \colhead{(arcsec)} & \colhead{} & \colhead{}
	}
	\startdata
	XMM/EPIC & 0821810401 & Schellenberger & Feb 19, 2019 & 26 & 10 & Medium & Full Frame \\
	Chandra/ACIS-I & 21065 & Ettori & Sept 29, 2018 & 10 & 0.5 & - & VFAINT 
	\enddata
\end{deluxetable*}
The \textit{Chandra} snapshot observation (see Tab. \ref{tab:xraydata}) was performed for the \textit{XMM-Newton} Cluster Heritage program  (\citealp{Arnaud2021-nz}) to find the cluster center for alignment of the upcoming \textit{XMM-Newton} observation. This \textit{Chandra} snapshot also provides a good reference of the X-ray point sources in the field.
We used the default Chandra data reduction using the CIAO package (\citealp{Fruscione2006-wt}) version 4.10 and CALDB 4.8.1, and followed the procedure described in \cite{Schellenberger2017-mc}. 
Point sources were detected using the \verb|wavdetect| task, and only point sources with a significance of at least 3 are shown (see green circles in Fig. \ref{fig:adative}).

Our \textit{XMM-Newton} EPIC observation (see Tab. \ref{tab:xraydata}) was reduced by the default tasks \verb|emchain| and \verb|epchain| within the SAS software package (version 15.0.0). CCDs 3 and 6 of MOS1 were excluded since they suffered from damage of hits by micro meteorites. No CCDs in an anomalous state were found.
For MOS data we included all events with \verb|PATTERN <= 12| and used the \verb|#XMMEA_EM| flagging, while for PN data we included only \verb|PATTERN <=4| and \verb|FLAG==0|. The PN data were corrected for out-of-time events. 
Further reduction steps followed the ESAS cookbook\footnote{https://heasarc.gsfc.nasa.gov/docs/xmm/esas/cookbook/xmm-esas.html} default analyis, including a lightcurve cleaning with \verb|mos-filter| and \verb|pn-filter|. We did not detect periods of high count rates to be removed and the full observation was used.
Spectral fitting to determine temperatures and abundances of the ICM was performed with Xspec (\citealp{Arnaud1996-uy}) following the procedure described in \cite{Lovisari2019-nh}.

\subsection{uGMRT radio data}
\begin{deluxetable*}{ccccccccc}
	\tablecaption{Summary of the radio GMRT observations \label{tab:radiodata}}
	\tablehead{
		\colhead{Project} & \colhead{Frequency} & \colhead{Bandwidth} & \colhead{Date} & \colhead{Time} & \colhead{FWHM} & \colhead{r.m.s} & \colhead{\#Ant} & \colhead{Calibrator} \\
		\colhead{}  &  \colhead{(MHz)} & \colhead{(MHz)} & \colhead{} & \colhead{(min)} & \colhead{(\si{\arcsec} $\times$ \si{\arcsec})} & \colhead{($\si{\mu Jy/beam}$)}& \colhead{} & \colhead{Bandpass/Phase}  \\
		\colhead{(1)}  &  \colhead{(2)} & \colhead{(3)} & \colhead{(4)} & \colhead{(5)} & \colhead{(6)} & \colhead{(7)}& \colhead{(8)} & \colhead{(9)}
	}
	\startdata
    38\_028 & 145 & 50 & June 16, 2020 & 210 & $18\times 15$ & 2000 & 27 & 3C286/1609+266\\
	38\_028 & 221 & 70 & June 16, 2020 & 210 & $12\times 10$ & 500 & 27 & 3C286/1609+266\\
	37\_018 & 383 &  200 & Nov 18, 2019 & 132  & $6.5\times 5.5$ & 43 & 26 & 3C286/3C298\\
	38\_028 & 650 & 188 & June 13, 2020 & 184 & $5.0\times 3.3$ & 13 & 29 & 3C286/3C298
	\enddata
	\tablecomments{Column 1: project code. Columns 2--5: observing frequency, bandwidth, date and on source time. Column 6: full-width half maximum (FWHM) of the array (obtained for ROBUST=0 in wsclean). Column 7: image r.m.s. level ($1\sigma$). Column 8: Number of active antennae. Column 9: Flux/bandpass and phase calibrators used.}
\end{deluxetable*}

The recently upgraded Giant Metrewave Radio telescope (uGMRT) provides new capabilities through  the four wideband receivers. We have observed Abell~2108 with uGMRT at band 2, 3, and 4 (see Tab. \ref{tab:radiodata}).
Band 3, $\SIrange{300}{500}{MHz}$, was recorded with the GWB backend using 4096 channels, 5.37s integration time, and full Stokes mode. The hardware radio frequency interference (RFI) filter\footnote{\url{http://www.ncra.tifr.res.in/ncra/gmrt/gmrt-users/online-rfi-filtering}} was turned on for this observation.
The data were split into 6 parts with a bandwidth of $\SI{33}{MHz}$ each, which were processed with the SPAM pipeline individually (\citealp{Intema2009-cs}). The noise level of the highest frequency part (centered at $\SI{482}{MHz}$) was about three times higher than the average noise in the five other parts ($\SI{100}{\mu Jy\,beam^{-1}}$), and so the highest frequency part was excluded from further analysis, causing the effective frequency of this band 3 observation to be $\SI{383}{MHz}$.

The follow-up observations in band 4, $\SIrange{550}{750}{MHz}$, and band 2, $\SIrange{100}{300}{MHz}$, were taken about 7 months after the initial band 3 detection with the same number of channels and integration time. The original scheduling could not be executed owing to the impact of cyclone Nisarga, which caused damage to the GMRT in early June 2020. With the excellent response and help of the GMRT staff our observations were rescheduled soon after with only a few antennas missing. The hardware RFI filtering was not available for these two observations. 

The GMRT Operations Team reported recently that continuum wideband observations taken in the last few years can be affected by systematics in the GWB correlator that give rise to a higher flux in the central square atnennas (baselines below 1\,km), most prominently in band 5\footnote{\url{http://indrayani.ncra.tifr.res.in/~secr-ops/sch/c39webfiles/central-square_23nov.txt}}. 
We tested whether our observations are affected by comparing the average phase calibrator amplitude of short baselines between GWB and the unaffected GSB correlator (which was recorded simultaneously for all our observations). We find no difference ($<0.5\%$) for band 3 and band 4, and for band 2 the effect of the systematics is expected to be negligible.

After excluding some channels at the edge of band 4, the observation was split into 4 parts ($\SI{47}{MHz}$ bandwidth). The noise levels of the individual parts were between 18 and $\SI{28}{\mu Jy\,beam^{-1}}$, and all 4 subsets were used for the combined analysis. Band 2 includes the old $\SI{150}{MHz}$ and $\SI{235}{MHz}$ receivers, which results in a non-continuous bandshape. The lower frequency part of this bandpass, $\SIrange{120}{170}{MHz}$, was split into two sub-parts with $\SI{25}{MHz}$ bandwidth, and the higher frequency part, $\SIrange{185}{255}{MHz}$, was split into two sub-parts with $\SI{35}{MHz}$ bandwidth.  We reconstructed images for the two subbands of uGMRT band 2 separately, reaching noise levels of $\SI{2}{mJy\,beam^{-1}}$ at 145\,MHz, and at $\SI{0.5}{mJy\,beam^{-1}}$ 221\,MHz.

Combined imaging of the GMRT observations was done with wsclean (\citealp{Offringa2014-bw}, version 2.10). To compensate for the widefield imaging we used the w-algorithm implemented in wsclean and increase the suggested number of w-planes by a factor of 5 for increased accuracy. The robust parameter of the Briggs weighting scheme was set to 0.5, except where noted otherwise, and the multiscale deconvolution algorithm was used. Wideband effects were taken into account by parameterizing the spectral behavior with a third order polynomial. After an initial run we used pyBDSF (\citealp{Mohan2015-hz}) on the preliminary images to detect structures ($6\sigma$ significance), which are masked for the clean algorithm in a final run of wsclean.

All final images have been primary beam corrected following the GMRT guidelines\footnote{\url{http://www.ncra.tifr.res.in/ncra/gmrt/gmrt-users/observing-help/ugmrt-primary-beam-shape}}. Since Band 2 has no updated primary beam shapes, yet, we used the primary beam parameterization of the pre-upgrade 235\,MHz band\footnote{\url{http://www.ncra.tifr.res.in:8081/~ngk/primarybeam/beam.html}}.

\section{Results}
\label{ch:results}

\begin{figure*}[htbp]
    \centering
    \includegraphics[width=0.91\textwidth]{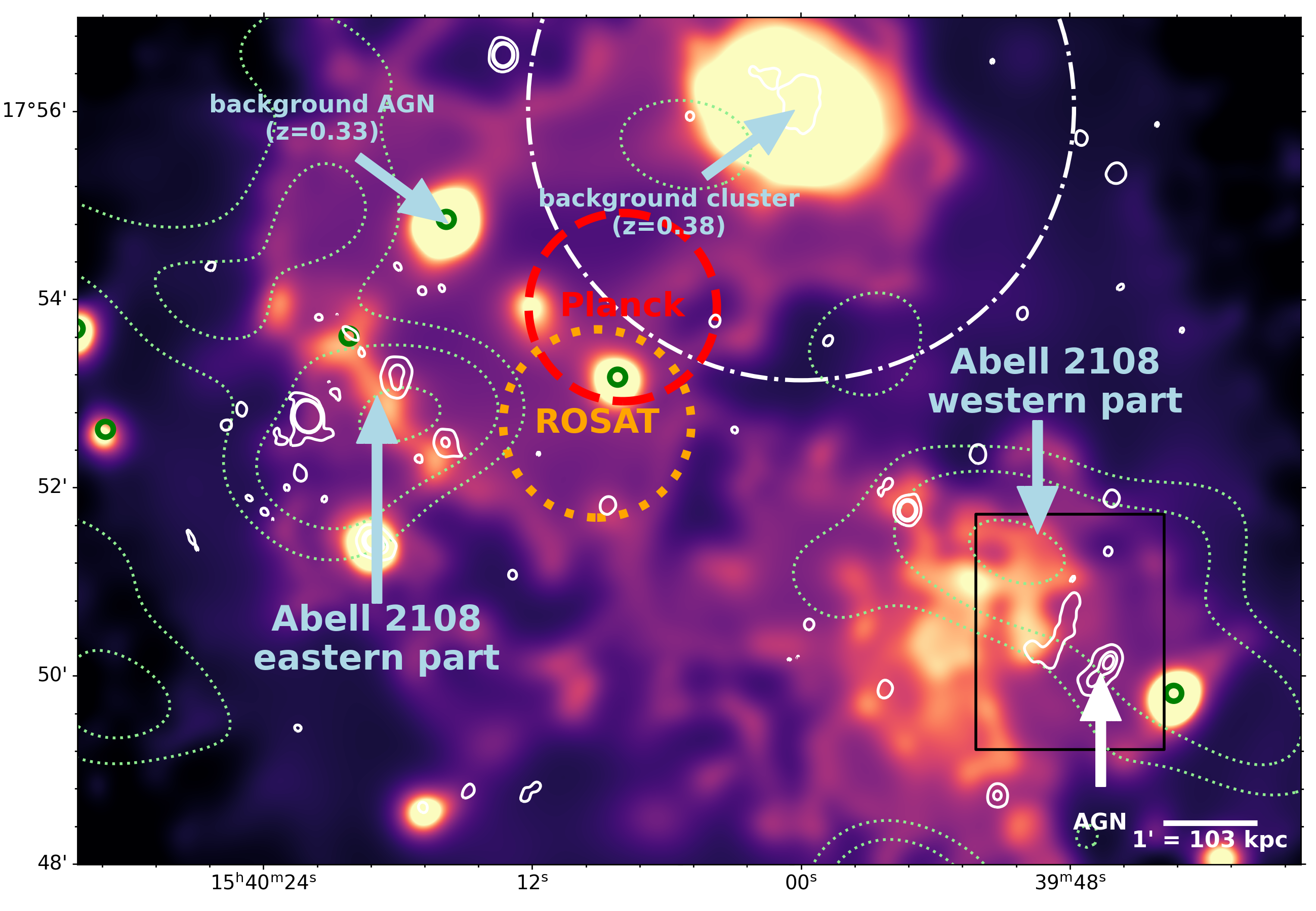}
    \caption{Adaptively smoothed, exposure corrected and background subtracted XMM-Newton image (0.4-1.6\,keV band). Green circles mark Chandra detected point sources. The dot-dashed white circle illustrates $R_{500}$ of the background cluster. The white contours show uGMRT Band 3 (383\,MHz) radio data at 5, 50 and 100 $\sigma$ level. Dotted contours reflect the member galaxy density using SDSS. The black rectangle shows the close-up region in Fig. \ref{fig:radio_relic}. The orange and red circle illustrate the ROSAT and Planck detection, respectively.}
    \label{fig:adative}
\end{figure*}

Abell~2108 was classified as a low richness cluster by \cite{Abell1958-dx}, and is part of the ROSAT Brightest Cluster Survey (BCS, \citealp{Ebeling1998-za}). 
Optical redshifts clearly place the cluster in the local universe ($z=0.092$, \citealp{Sarazin1982-az}). 
The homogenized ROSAT X-ray luminosity of the field is rather low ($L_X=\SI{1e44}{erg/s}$ in the $\SIrange{0.1}{2.4}{keV}$ band, \citealp{Piffaretti2011-ay}), and luminosity-mass scaling relations place it at the lower mass end of clusters ($\SI{2.0(3)e14}{M_\odot}$, \citealp{Schellenberger2017-mc}). 
However, the strong SZ pressure signal, and the SZ signal to noise ratio of $6.9$ in the Planck survey (\citealp{Planck_Collaboration2016-jb}, PSZ2 G028.63+50.15) indicates a higher mass of about $\SI{3.2(3)e14}{M_\odot}$ (using the M-Y$_{\rm{SZ}}$ relation by \citealp{Arnaud2010-ct}).
The locations of ROSAT and Planck detections are offset by $\SI{1.3}{arcmin}$.

The large mass discrepancy between the X-ray and SZ intrigued us to study this object in more detail, as we suspected a shock heated ICM from a recent major merger in a low mass system. We first describe the system based on our follow-up observations with XMM-Newton and the Chandra snapshot, and then summarize our findings from the SZ, optical and radio data.

\subsection{X-ray perspective}
Our X-ray observations show that the field can be subdivided in three parts: A western and eastern part of Abell~2108, and a background cluster to the north (Fig. \ref{fig:adative}).

\subsubsection{The background cluster} 
The X-ray brightest feature in the field is located in the north with a flux in the $\SIrange{0.1}{2.4}{keV}$ band of $\SI{1.3e-12}{erg\,s^{-1}\,cm^{-2}}$ as estimated from the \textit{XMM-Newton} data. It is a cluster of galaxies at a redshift of 0.38 (WHL J154000.1+175608, confirmed with SDSS spectroscopic redshifts, \citealp{Wen2010-hl}, and with the X-ray spectral fitting).
The X-ray peak J2000 coordinates are R.A.$15^{\rm{h}}40^{\rm{m}}00^{\rm{s}}$ Decl.$+17^{\rm d}56^{\rm m}05^{\rm s}$.

From about $\num{20000}$ counts in our XMM-Newton observation we can estimate a reliable temperature and abundance ($\SI{4.34(15)}{keV}$, $\SI{0.47(2)}{Z_\odot}$) within half $R_{500} = \SI{1.5}{arcmin}$. 
Utilizing the mass-temperature scaling relation by \cite{Lovisari2020-gn}, we derive $M_{500} = \SI{3.3(3)e14}{M_\odot}$. The dot-dashed circle in Fig. \ref{fig:adative} shows $R_{500}$ around the background cluster. However, the luminosity-mass scaling relation by \cite{Lovisari2020-gn} gives a significantly higher mass of $\SI{5.6(5)e14}{M_\odot}$. 
The residual ROSAT flux, after subtracting the background cluster, converts to a mass of Abell~2108 (East+West) of $\SI{1.8(4)e14}{M_\odot}$.

\subsubsection{Abell~2108 West} 
\label{ch:results_west}
The western part of Abell~2108, with the X-ray peak located at R.A.$15^{\rm{h}}39^{\rm{m}}52{\rm{s}}$ Decl.$+17^{\rm d}56^{\rm m}05^{\rm s}$, has an average temperature of $\SI{3.0(1)}{keV}$, and a low metallicity ($\SI{0.2(1)}{Z_\odot}$), despite it being the brightest peak in the X-rays of Abell~2108. 
This temperature corresponds to a cluster mass of ${\SI{1.81(17)e14}{M_\odot}}$ using the mass-temperature (M-T) scaling relation from \cite{Lovisari2020-gn}.

\begin{figure*}
    \includegraphics[width=0.9\textwidth]{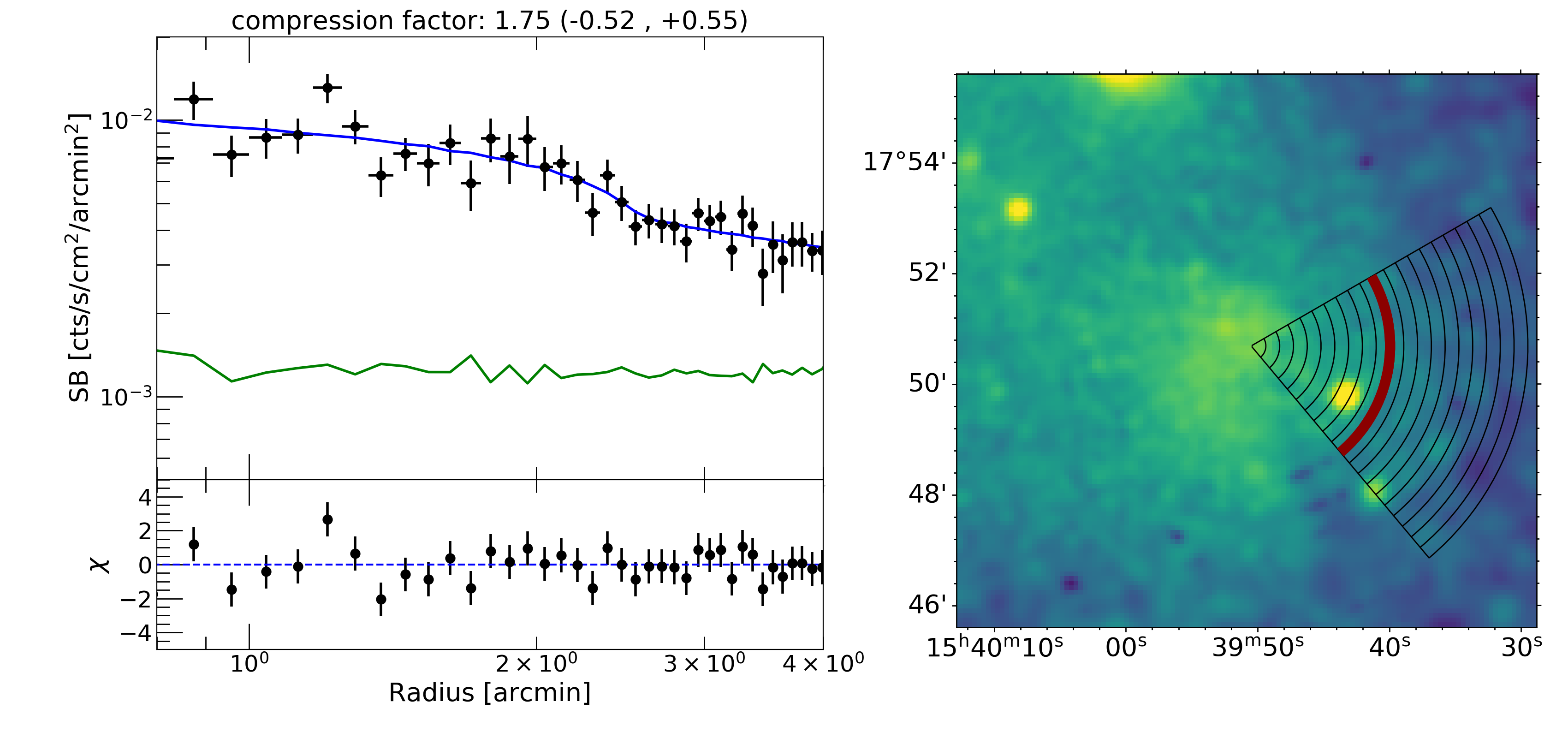}
    \caption{\textit{Left:} $\SIrange{0.4}{4}{keV}$ X-ray surface brightness profile along the western direction of the western cluster part. Model fit (blue line) is performed with \textit{PROFFIT}. The green line shows the particle background level. \textit{Right:} Location of the extraction region and the discontinuity (dark red line).}
    \label{fig:sbr_edge}
\end{figure*}

Subdividing the western clump into smaller pieces gives roughly the same ICM parameters. At the western border of this clump we find indications for a surface brightness edge.
We extracted the circular surface brightness profile in small sectors in the direction of the temperature discontinuity, and used the software package \verb|PROFFIT| (\citealp{Eckert2020-xx}) to fit a broken powerlaw convolved with the XMM-Newton PSF. 
The profile was centered on the center of the western part (see Fig. \ref{fig:sbr_edge}) and uses logarithmically sized bins with a minimum of 5\,arcsec width for each bin of the sector. This results in relative uncertainties of the surface brightness in each sector between 10 and 20\%. 
We find a discontinuity west of the main subcluster. Figure \ref{fig:sbr_edge} shows the profile and the model fit for the selected sector. 
We tested elliptical surface brightness profiles, and several opening angles of the sector, which typically detect no or a weaker edge. 
Between the main subcluster and the surface brightness edge we find a  higher ICM temperature ($\SI{3.3(2)}{keV}$), while going even further to the west the temperature seems to drop rapidly to about $1.84^{+0.16}_{-0.12}\,\si{keV}$. 
This is a strong indication for a shock front, which are rarely observed among poor clusters (e.g., \citealp{Russell2014-zj}). 
Based on the Rankine–Hugoniot criterion (\citealp{Landau1959-li}) we can derive a Mach number for this shock from the X-ray temperature jump: 
We find $\mathcal{M}_\text{X} = \num{1.8(2)}$, which is a weak shock, as expected for a low mass cluster merger. We note that the PSF of XMM-Newton will affect the ability to detect sharp temperature discontinuities, therefore the temperature jump is likely higher. 
This is in agreement with the outer surface brightness discontinuity, from which we derive a Mach number of $\mathcal{M}_\text{SBR} = 1.5^{+0.5}_{-0.4}$ based on the density jump.

\subsubsection{Abell~2108 East} 
Abell~2108 East (Fig. \ref{fig:adative}) can be split into a linear, brighter stripe (NE-SW direction with the X-ray peak located at R.A.$15^{\rm{h}}40^{\rm{m}}19^{\rm{s}}$ Decl.$+17^{\rm d}52^{\rm m}53^{\rm s}$), and larger scale diffuse emission with an average temperature $\SI{2.9(2)}{keV}$ and a metallicity $\num{0.5(2)}$. Using again the M-T relation of \cite{Lovisari2020-gn} we find a mass of $\SI{1.7(2)e14}{M_\odot}$. 

The bright stripe has only one point source in the Chandra snapshot at the eastern end, and is not correlated with the XMM-Newton EPIC chip gaps or other detector structures. 
We therefore have confidence that it is a real structure, not an artifact.
The width of the stripe corresponds roughly to the XMM-Newton PSF, so it may not be resolved by XMM-Newton, but is also too faint to be detected in the short Chandra observation. The temperature of the stripe is similar to the overall eastern clump. 

We also see two bright clumps of emission with significantly cooler temperatures, both of which contain a Chandra point source. The position of the northern one coincides with a quasar at higher redshift, but the fainter clump could resemble a cool core or galaxy halo. Neither are located in the brightest parts of the ICM, or the densest parts of the galaxy population, as would be expected for a cool core. 

The region between the eastern and western clumps is very faint, and due to the XMM-Newton PSF and its high and variable background it is not clear if there is any excess emission between these two components of the merger system.

\begin{deluxetable}{ccccc}
	\tablecaption{Summary masses derived from X-ray scaling relations ($T-M$ and $L-M$ by \citealp{Lovisari2020-gn}) and the SZ scaling relation ($Y_\text{SZ}-M$ by \citealp{Arnaud2010-ct}). \label{tab:masses}}
	\tablehead{
		\colhead{} & \colhead{z} & \colhead{$Y_\text{SZ}-M$} & \colhead{$T-M$} & \colhead{$L-M$} \\
		\colhead{} & \colhead{} & \colhead{[$\SI{e14}{M_\odot}$]} & \colhead{[$\SI{e14}{M_\odot}$]} & \colhead{[$\SI{e14}{M_\odot}$]} }
	\startdata
	whole field & 0.09 & $3.21_{-0.57}^{+0.53}$ & - & $\num{2.2(4)}$ \\
	\hline
	Abell~2108-E & 0.09 & $\num{1.4(2)}$ & $\num{1.7(2)}$ & \multirow{2}{*}{$\num{1.8(4)}$} \\
	Abell~2108-W & 0.09 & $\num{1.7(2)}$ & $\num{1.8(2)}$ & \\
	Bkg. cluster & 0.38 & $4.5_{-0.9}^{+1.1}$ & $\num{3.3(3)}$ & $\num{5.6(5)}$\\
	Bkg. cluster & 0.09 & $\num{1.3(3)}$ & - & $\num{1.1(3)}$
	\enddata
	\tablecomments{Row 1: Masses from SZ/X-ray flux assuming background cluster is also at Abell~2108 redshift. Row 2-4: Three individual cluster components in the field. Row 5: Background cluster masses derived for redshift of Abell~2108.}
\end{deluxetable}
\subsection{Planck Sunyaev-Zeldovich masses}
As we have seen in the previous section, the field of Abell~2108 can be separated into three components, Abell~2108 east and west, and a background cluster. The three components are too close to be separated by Planck, however it is possible to extract SZ model normalizations from Planck for the three components with some X-ray priors. We use three cluster components following the \cite{Arnaud2010-ct} pressure profile as a template, and center each on the X-ray peaks with SZ fluxes corresponding to the X-ray luminosity. We then filter the Planck maps with Matched Multifilters using the templates as inputs, and obtain SZ normalizations for the three components in a two-parameter fit (background cluster and Abell~2108 east+west). We find $Y_{500}^{\rm BG} = \SI{5.23(206)e-4}{arcmin^2}$ for the background cluster, $Y_{500}^{\rm A2108W} = \SI{8.15(161)e-4}{arcmin^2}$ for Abell~2108 west, and $Y_{500}^{\rm A2108E} = \SI{5.59(110)e-4}{arcmin^2}$ for Abell~2108 east, and converted these SZ signals to the masses shown in Tab. \ref{tab:masses} ($Y_\text{SZ}-M$ column) using Eq. 25 from \cite{Arnaud2010-ct}.
Overall, we find good agreement with the X-ray temperature derived masses of the individual components. Comparing the sum of the SZ mass of the Abell~2108 components and the mass from the Abell~2108 luminosity (east + west), we find with 99.8\% confidence that this object is X-ray underluminous.

\subsection{Optical perspective}
\label{ch:optical}
Looking at the distribution of galaxies in a merging cluster can help to understand the merging scenario. We use the SDSS catalog of \cite{Alam2015-yc} to select 412 member galaxies from the red sequence in the color-magnitude diagram, and verify our selection with the spectroscopically confirmed member galaxies in the same diagram. We show the member galaxy density  in Fig. \ref{fig:adative} as light green, dashed contours. We see a similar distribution as in the X-rays: two sub-clusters of galaxies, where the eastern one coincides with the X-ray bright stripe, and the western one is shifted slightly to the north of the X-ray emission. 

76 of the galaxies have SDSS spectroscopic redshifts, which show the galaxies in the western part to be slightly blueshifted, with a line-of-sight velocity difference to the eastern ones of $\SI{603(180)}{km\,s^{-1}}$.
From the two distinct galaxy distributions and the disturbed X-ray morphologies of the parts, one can conclude that there is a major merger ongoing, likely after core passage. 
The Mach numbers derived in section \ref{ch:results_west} predict a merger velocity $\SI{1090(150)}{km\,s^{-1}}$. From the measured line-of-sight velocity we derive an inclination angle of $34^\circ\,\pm\,14^{\circ}$ between the merger axis and the plane of the sky.
This means that small projection effects lead to an underestimate of the X-ray Mach number (e.g., \citealp{Machado2013-em}). From the (projected) location of the X-ray edge (Section \ref{ch:results_west}) we conclude that the merger happened roughly 0.7\,Gyr ago. This is relatively long, but not exceptional (see \citealp{Golovich2019-zd}). We note that shocks may accelerate after initial launch (e.g., as seen in simulations by \citealp{Ha2018-ou}) making this timescale a lower limit. 

\cite{Wen2013-rf} use SDSS data to compute dynamical state and cluster richness values from the optical survey data. From the provided radius ($r_{200}$) we compute the cluster mass $M_{500}$ using the simple relation $r_{500} = 2/3 r_{200}$ (\citealp{Shimizu2003-nv}), and find $\SI{2.6e14}{M_\odot}$ for Abell~2108, and $\SI{3.3e14}{M_\odot}$ for the background cluster. The mass of Abell~2108 is in broad agreement with what has been found from X-ray and SZ. However, the background cluster was expected to be more massive based on its X-ray luminosity or SZ signal. 

\begin{figure*}[t]
    \centering
    \includegraphics[width=0.99\textwidth]{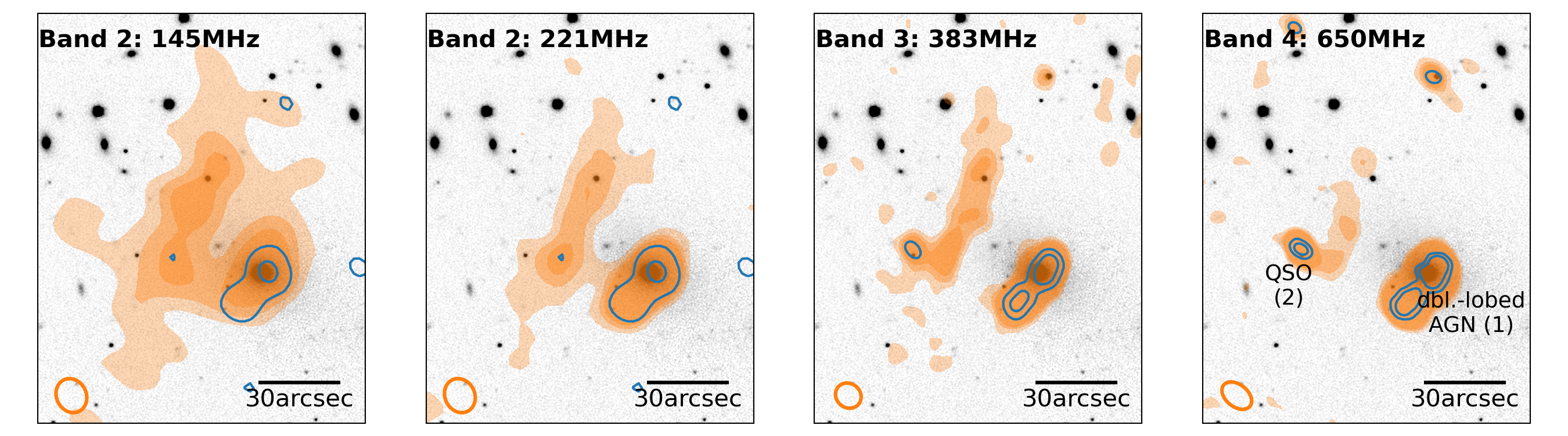}
    \caption{SDSS r-band image with filled orange contours of the extended radio emission (2, 4, 6$\sigma$). Blue contours showing the compact radio emission (5, 25 $\sigma$). Each panel shows a separate uGMRT band, with band 2 split into the two subbands described in the text. The beam size is shown in the lower left corner.}
    \label{fig:radio_relic}
\end{figure*}
\subsection{Radio emission}
\label{ch:results_radio}
Our GMRT observations of Abell~2108 in four bands provide a comprehensive view in the radio regime. The rapid decrease in surface brightness, and the temperature drop in the XMM-Newton spectral analysis gives strong hints for a shock front. Our radio observations enable us to look closer at the interesting location to search for  relativistic particles producing synchrotron radiation. In Fig. \ref{fig:radio_relic} we show the resulting images in the three bands and distinguish between compact emission (detected by the longest baselines and shown in blue contours in Fig. \ref{fig:radio_relic}), and extended emission (from all the baselines shown as orange filled contours in Fig. \ref{fig:radio_relic}).

The brightest radio source in this region is a double lobed radio AGN (1 in Fig. \ref{fig:radio_relic}) associated with the galaxy \verb|WISEA J153946.45+175007.9|, which has a flux of $\SI{11.0(1)}{mJy}$ at $\SI{650}{MHz}$, and a consistent spectral index\footnote{We define the spectral index $\alpha$ as $S_\nu \propto \nu^\alpha  $, where $S_\nu$ is the flux density at frequency nu.} of $\alpha = -0.7$ between 188 and 650\,MHz. SDSS spectroscopic redshift (0.0898) of the galaxy shows it to be a member of Abell~2108. \cite{Lin2018-jt} determine a stellar mass of the host galaxy of this AGN of $\SI{9.3e10}{M_\odot}$. We do not detect any X-ray emission associated with the galaxy or AGN.

More compact radio emission is detected east of the AGN region (marked by 2 in Fig. \ref{fig:radio_relic}), arising from a high redshift ($z > 4$) quasar (\citealp{Richards2015-ae}). This source has an inverted radio spectrum with a spectral index $\alpha \approx +0.8$, which is in good agreement with the flux density quoted in the FIRST survey at $\SI{1.4}{GHz}$. The inverted spectrum predicts a faint flux density for band 2 ($\SI{0.2}{mJy}$), accounts for our band 2 non-detection.

East of \verb|WISEA J153946.45+175007.9| (Fig. \ref{fig:radio_relic}) we find very extended and elongated radio emission, especially in band 2 and 3. The shape in band 2 is very diffuse and extended primarily in N-S direction. The maximum size is $200\times \SI{60}{kpc}$.  
Band 3 has a similar extent in the N-S direction, but is more confined in E-W, and bent to the east at the southern end. Less emission is detected in band 4 despite the low noise, indicating a very steep spectrum. 
In the higher bands, we find no radio emission connecting the western AGN (1 in Fig. \ref{fig:radio_relic}) and the extended emission, therefore we think it is unlikely that the extended emission is related to this galaxy. However, we cannot exclude this possibility.

Extended emission is clearly detected, with a peak flux to noise ratio of 7, 11, 5 for band 2 (combined), 3, and 4, respectively. 
To estimate the flux of the extended emission, we measure the total flux in the field and subtract the contributions of the quasar and double-lobed AGN. The residual fluxes are $\SI{100(9)}{mJy}$ at 145\,MHz, $\SI{30.4(25)}{mJy}$  at 221\,MHz, $\SI{6.9(13)}{mJy}$ at 383\,MHz, and $\SI{1.2(7)}{mJy}$ at 650\,MHz. The uncertainties include a systematic uncertainty of 5\% on the measured total flux.

\begin{figure*}
    \centering
    \includegraphics[width=0.49\textwidth]{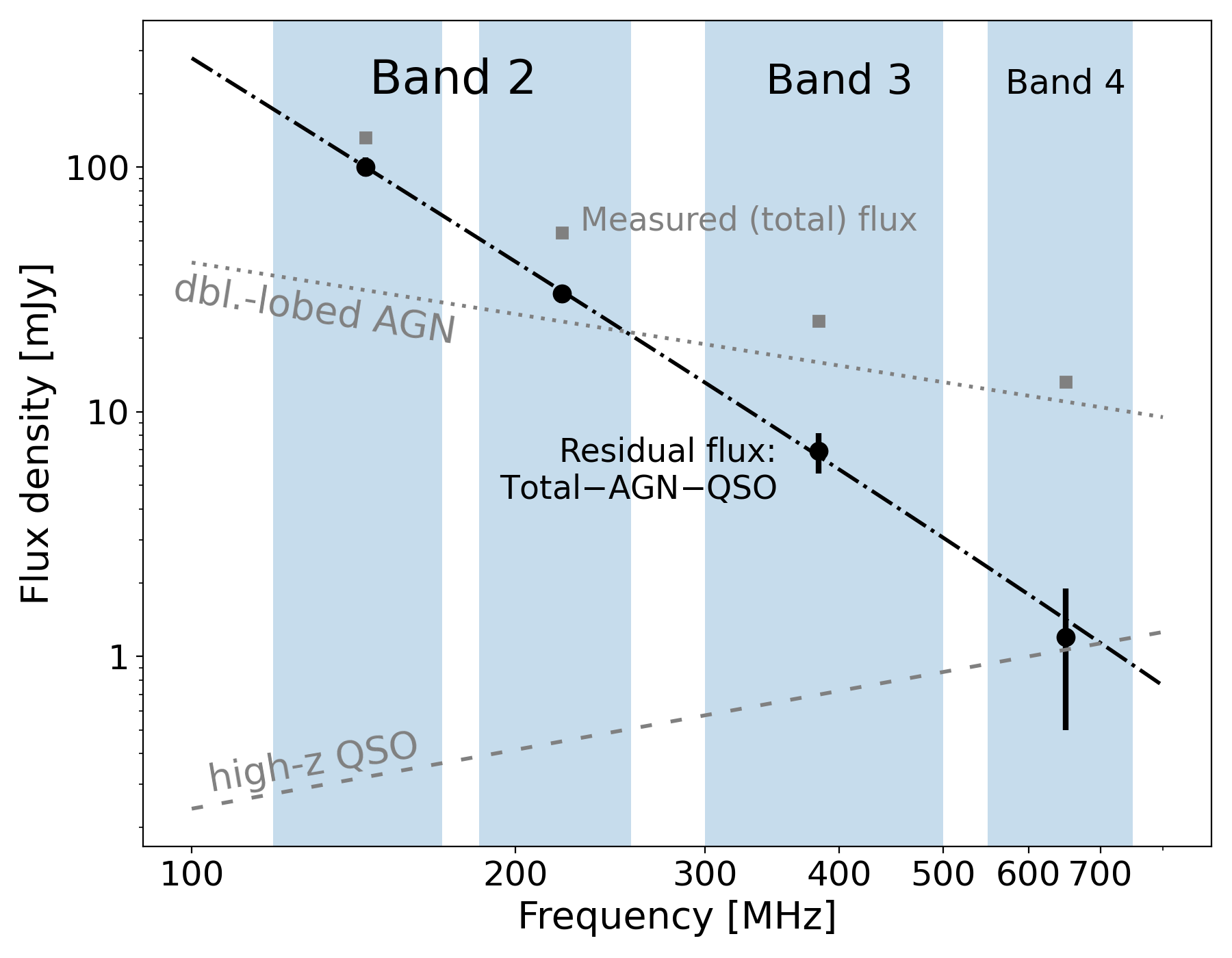}
    \includegraphics[width=0.49\textwidth]{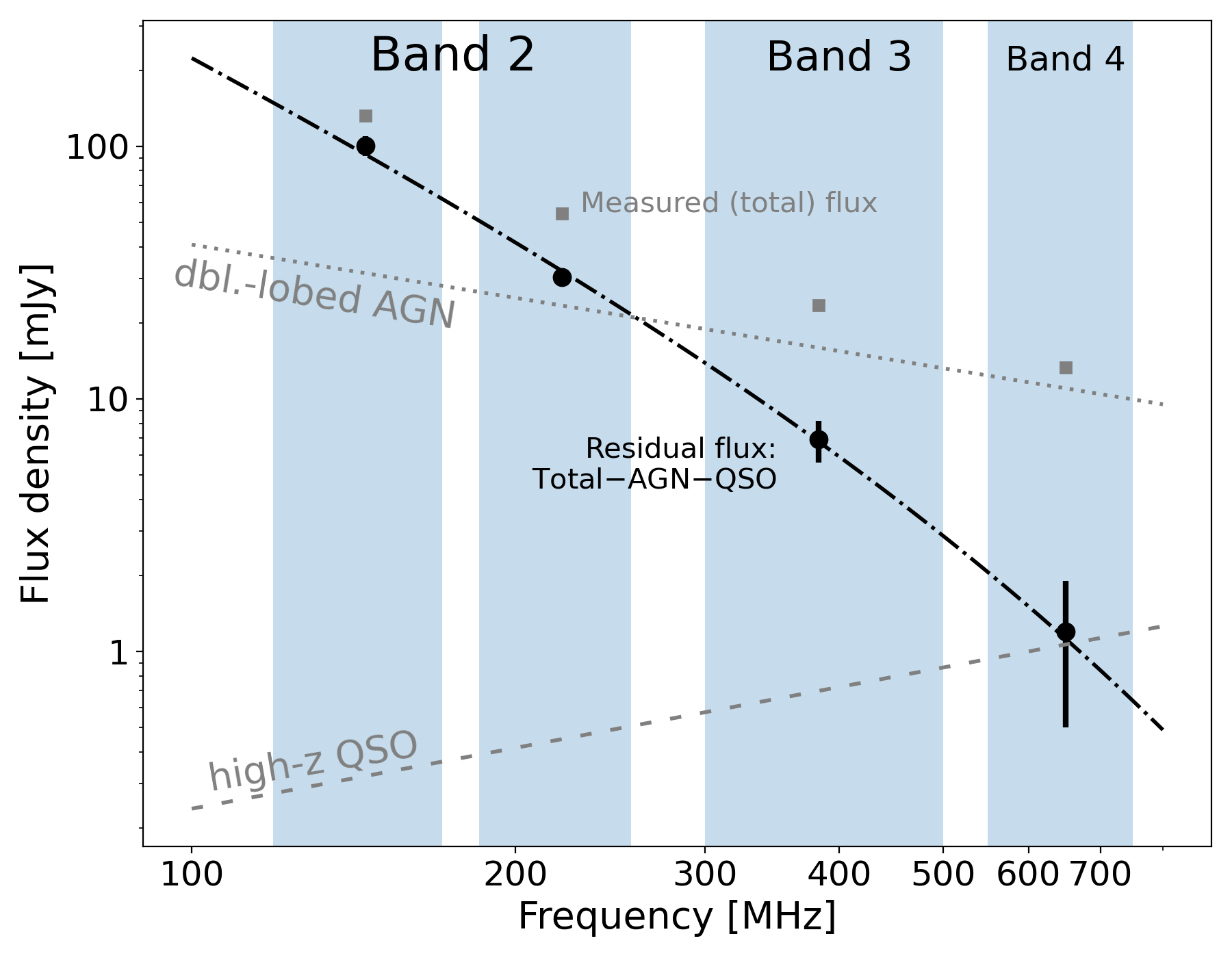}
    \caption{Radio spectrum of the extended emission in Abell\,2108. The total fluxes are shown as gray boxes, from which contributions of the double-lobed AGN (dotted line) and the quasar (dashed line) are subtracted to yield the residual fluxes (black circles). The left panel shows the fit (black dotted-dashed line) without priors, while in the right panel  a prior is set on the injection index based on the X-ray derived Mach number.}
    \label{fig:radio_spectrum}
\end{figure*}

The radio residual flux spectrum is consistent with a steep powerlaw (Fig. \ref{fig:radio_spectrum} left), a prediction of direct diffusive shock acceleration (direct DSA), where shock-energized thermal electrons emit synchrotron radiation in the presence of the weak ($\sim \si{\mu G}$) magnetic fields within the cluster. 
We note that due to the offset between the radio emission and the X-ray inferred shock location we exclude this possibility, and provide a more detailed discussion in section \ref{ch:discussion_shock}. 
Instead, our preferred model for the spectrum of the radio emission includes an aging term to describe synchrotron cooling losses. We fit the residual flux spectrum with a powerlaw including a high-frequency exponential cut-off (Fig. \ref{fig:radio_spectrum} right), $S_\nu \propto \nu^\alpha e^{-\nu / \nu_{\rm break}}$, resembling an aged synchrotron spectrum (MJP) as described in \cite{Slee2001-jw}. We set $-2$ as a prior for the observed spectral index $\alpha$,
which corresponds to the radio spectral index in case of re-acceleration by a Mach $1.6$ shock, and is significantly shallower than the overall spectrum.
The fit provides a break at $\nu_{\rm break} = 361^{+209}_{-103}\,\si{MHz}$. However, the error bars are large due to the uncertainty on the band 4 data point, which is crucial to determine any break in the spectrum. 

\section{Discussion}
\label{ch:discussion}
\begin{figure*}[htbp]
    \centering
    \includegraphics[width=0.91\textwidth]{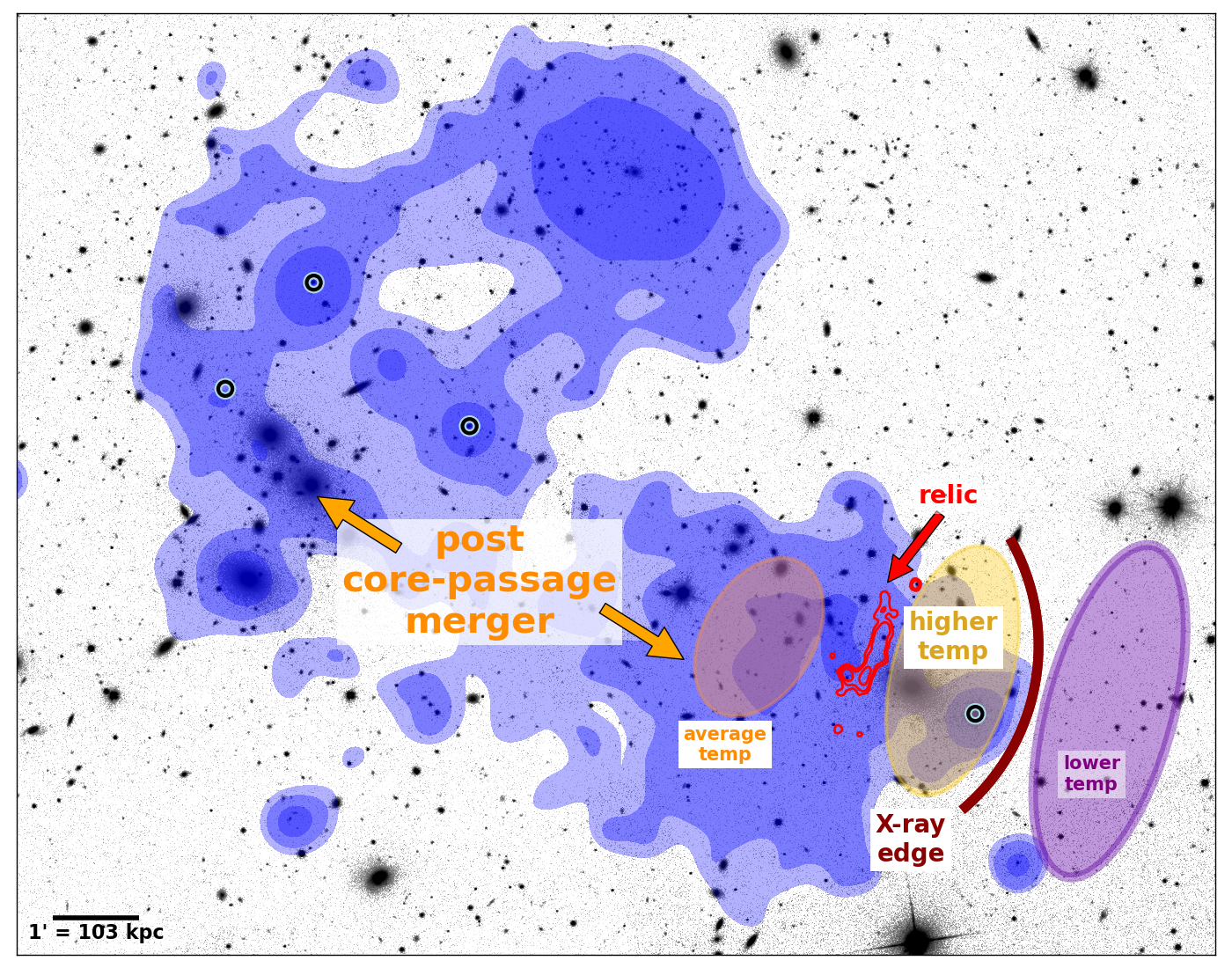}
    \caption{SDSS r-band image to illustrate the merger scenario in Abell~2108: The X-ray emission is shown by blue filled contours, and the Chandra detected X-ray point sources by small black circles. Filled ellipses/circles demonstrate regions for X-ray spectra to determine the ICM temperature, and the X-ray surface brightness edge is shown in dark red. Red contours show the radio relic in uGMRT band 3.}
    \label{fig:scenario}
\end{figure*}
It is evident from the X-ray observations that Abell~2108 is a merging system whose eastern and western subclusters both exhibit a disturbed morphology. 
Based on the X-ray and radio observations we have strong indications that we have found a shock front and radio relic in the western part. The evidence is summarized in Fig. \ref{fig:scenario}: The X-ray surface brightness profile shows the edges (dark red line in Fig. \ref{fig:scenario}) indicating a shock front between the region of elevated ICM temperature ($\SI{3.3}{keV}$, post-shock region in Fig. \ref{fig:scenario}), while out to the west the ICM temperature drops significantly ($\SI{1.8}{keV}$, pre-shock region in Fig. \ref{fig:scenario}). The location of this edge and temperature discontinuity is to the west of the western cluster part along the merger axis, where an axial shock front is usually expected (\citealp{Ha2018-ou}). 
The large amount of released gravitational energy from mergers not only heats the ICM, but also accelerates the population of non-thermal particles and amplifies magnetic fields (e.g., \citealp{Donnert2018-an}), leading to synchrotron radiation in the radio band. These radio relics are found in major mergers, usually of very high total mass.

\subsection{Aged radio emission from past shock re-acceleration}
\label{ch:discussion_shock}
We find extended radio emission in three uGMRT bands (the band 3 contours of the relic are shown in red in Fig. \ref{fig:scenario}). The shape resembles a radio relic located at the periphery of the cluster with a slightly bent morphology. Radio flux contributions from cluster galaxies and point sources have been taken into account when fitting the spectrum of the extended emission. The observed radio spectrum allows an exponential steepening at higher frequencies. 
The smaller extent of the emission in band 4 compared to the other bands may be caused by instrumental effects (flux losses caused by a non uniform coverage of the uv plane at short spacings in band 4), but it could also be a real effect, i.e., the spectral index is steeper in those regions. 
As discussed, e.g., by \cite{Giacintucci2008-dx} radio relics can be connected to disturbed clusters by shock acceleration, adiabatic compression, or shock re-acceleration. 
The re-acceleration of fossil, non-thermal electrons is likely if a source for this seed particle population can be identified. The proximity of the relic in Abell~2108 to the nearby AGN (projected separation $\sim \SI{50}{kpc}$) is interesting, since it raises the possibility that fossil relativistic particles from the AGN lobes form the seed population for the radio relic (e.g., \citealp{Van_Weeren2017-ej}). 
\cite{ZuHone2020-rp} discuss another possibility, where the seed population originates from fossil lobes of cluster AGNs disrupted by merger-driven gas motions, and their radio plasma has been mixed with the ICM. 

In the pure DSA model, where thermal electrons are energized by the shock to emit synchrotron radio emission, the shock front is expected to be coincident with the edge of the relic. However, the X-ray inferred shock location (from the surface brightness edge and temperature discontinuity) is further west of Abell~2108 West, which agrees with the general morphology of the merger, but is significantly offset from the location of the extended radio emission (Fig. \ref{fig:scenario}). 
Furthermore, DSA from the thermal pool is generally found to be unable to generate the observed radio emission in clusters because weak shocks (such as galaxy cluster merger shocks) have low acceleration efficiency (e.g., \citealp{Kang2013-nf}). 
For this reason, re-acceleration of pre-existing relativistic seed particles (via the same DSA physics) is often considered to explain the observed radio emission in relics (e.g., \citealp{Botteon2020-co}, and references therein). However, the spatial offset  in Abell~2108  might suggests that the shock is not currently re-energizing particles in the relic.

An offset between the shock front and the radio relic location was falsely identified in the Toothbrush cluster (\citealp{Ogrean2013-bi}) where a surface brightness edge was found in XMM-Newton data offset from the radio relic emission.
Deep Chandra observations found the X-ray shock front matching the location of the radio relic (\citealp{Van_Weeren2016-gt}), and small-scale substructure together with the poor spatial resolution of XMM-Newton was identified as the cause of the wrongly identified shock front from XMM-Newton data. 
In the case of Abell~2108 the angular offset between the radio relic and the X-ray discontinuity is twice as large as for the Toothbrush cluster. Therefore, it is unlikely that the offset is an artifact caused purely by the XMM-Newton PSF. 
With the shock front deviating from the radio relic location, we can exclude the possibility that particles are currently re-accelerated at the relic. 

Instead, we favor the scenario of a lack of seed electrons beyond the visible relic location. 
These pre-existing relativistic seed electrons are located in a cloud, and have been re-accelerated when the shock crossed the relic location in the past. After energizing the particles in this region, the shock front traveled to the current location (as traced by the X-ray data). 
Outside of this cloud there exists only thermal electrons, which cannot be accelerated efficiently (see references above), and the synchrotron emitting electrons lose some of their energy during the shock travel time after leaving the cloud. 
This means that the relic we see in Abell~2108 is left behind the shock and the synchrotron spectrum has aged. 
We tested this hypothesis by fitting the integrated radio spectrum with an MJP-like model, and derive a break frequency for the overall relic spectrum of $\nu_{\rm break} = 361^{+209}_{-103}\,\si{MHz}$. This assumes the injection spectral index to be $-2$, which corresponds to the radio spectral index in case of re-acceleration by a shock with Mach number $1.6$. 
Since the shock passes across the radio emission in a shorter timescale ($\sim \SI{50}{Myr}$) than the characteristic synchrotron cooling timescale of the re-energized non-thermal electrons ($>\SI{100}{Myr}$), the injection spectral index is equal to the observed spectral index at lowest frequencies before the break. 
For an average magnetic field strength of $\SI{3}{\mu G}$, which is derived assuming energy equipartition (Eq 25 in \citealp{Govoni2004-dx}, using $\xi=\num{7.76e-14}, \alpha=-2$, equal energy in relativistic protons and electrons, and assuming a prolate 3D shape of the relic), we find a spectral age of $\SI{150(30)}{Myr}$ based on the spectral break around 360\,MHz. This is comparable to the travel time of the shock front from the radio emission location to the observed X-ray discontinuity ($140^{+40}_{-80}\,\si{Myr}$) assuming a shock velocity based on the X-ray Mach number. Relaxing some of the assumptions for calculating the magnetic field strength yields a range of values from $1.5$ to $\SI{5}{\mu G}$, allowing aging timescales between 100 and $\SI{170}{Myr}$. 

Adiabatic compression of fossil non-thermal plasma by shocks (\citealp{Enslin2001-jg}), sometimes referred to as radio phoenices, provides a possibility to explain the radio features in Abell~2108, since radio phoenices typically show spectral steepening (which the band 4 data doesn't exclude). \cite{Kale2009-yx,Kale2012-ok} describe the different phases of radio synchrotron emission before and after shock-induced adiabatic compression. Due to the offset of the X-ray shock front, we can only consider the last ``fading'' phase for Abell~2108, characterized by a shift of the break toward lower energies in an already curved spectrum. The current depth of our higher frequency observations (uGMRT band 4) does not allow to clearly measure the spectral break, and deeper radio data are needed to demonstrate the possibility of adiabatic compression. 

Very steep spectrum radio remission without a bright point source could in principle also originate from remnant lobe of a radio galaxy that has been detached by ICM motion. However, there is no nearby member galaxy where a conclusive scenario could be constructed, with the possible exception of the double lobed radio source to the west (Fig. \ref{fig:radio_relic}). If the relic emission is a detached lobe of this AGN, it would have had to be transported toward the center of the western part of Abell~2108, in conflict with buoyant forces which should move it outward. Together with the orientation of the elongation in N-S direction, we think that this possibility is very unlikely. 

\subsection{Systematic uncertainties in tracing the shock}
The X-ray temperature discontinuity allows us to derive a Mach number that indicates a weak shock. From the velocities of the member galaxies we conclude an inclination angle of about $34^\circ$, close to the plane of the sky. 
A significant difference between X-ray and radio derived Mach numbers has been reported for many systems in the past (e.g., \citealp{Botteon2020-co}).
It is possible that viewing angles bias the X-ray derived Mach numbers and shocks are more complicated than in idealized simulations. 
Radio observations of relics trace the non-thermal component of cosmic rays while X-ray observations of shocks are sensitive to the kinetic energy (\citealp{Ha2018-ou,Hoeft2007-cv}). 
As pointed out by \cite{Hoang2017-pu} X-ray observations of shocks suffer from additional systematics due to inhomogeneities in the ICM and electron temperatures not tracing the ion population after shock heating. It is possible to address the latter point by deep observations of transition lines in the X-ray spectra. Inhomogeneities in the ICM will be difficult to quantify in a disturbed system like Abell~2108, where clumping might only be detected by future X-ray observatories. 
Fortunately, we do not expect the background cluster to complicate matters, since the relic is located so far away (roughly $2 R_{500}$ of the background cluster) that we do not expect any significant contaminating flux from its ICM.

\subsection{Biases in cluster mass measurements}
The detection algorithms used for the ROSAT and Planck surveys do not detect the three components in the Abell~2108 field as separate structures. However, the information from X-ray follow-up observations helped to constrain SZ properties of each component. If the background cluster was in the field, the SZ signal to noise would be around 5, removing it from the CHEX-MATE sample (\citealp{Arnaud2021-nz}).
\begin{figure*}[htbp]
    \centering
    \includegraphics[width=0.99\textwidth]{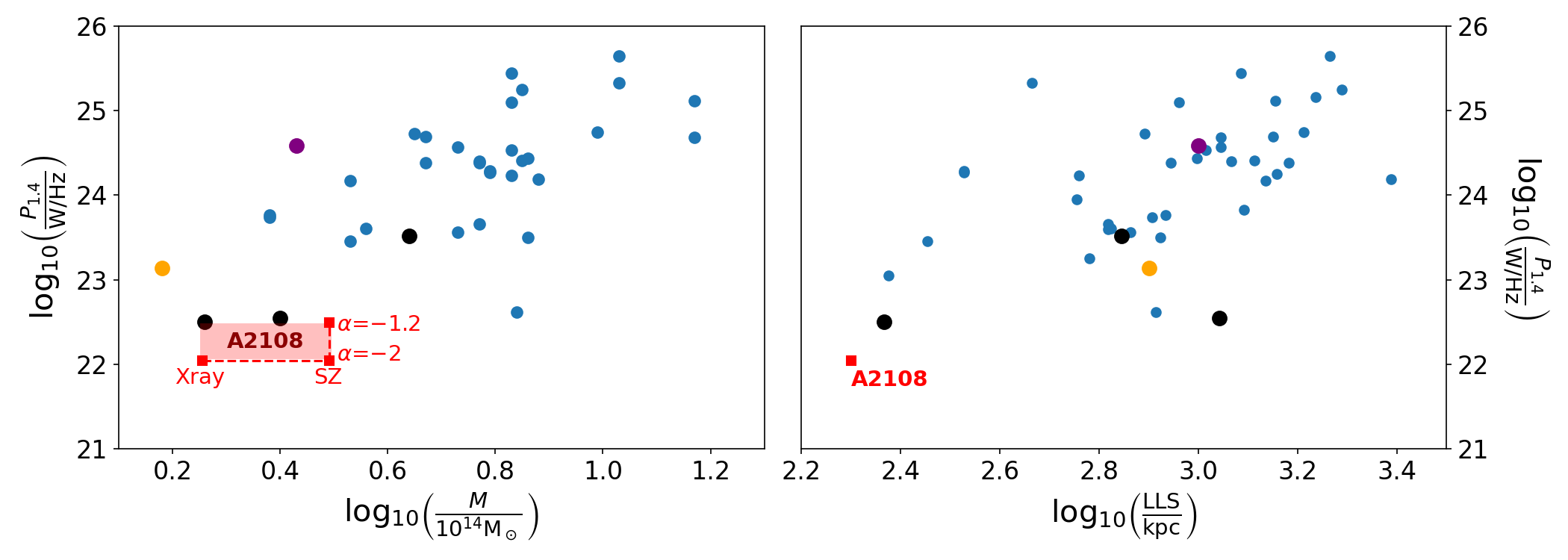}
    \caption{Radio power at 1.4\,GHz of know relics versus the host cluster mass (left), and the relic largest linear size (right). Relics are taken from literature (the large collection by \citealp{Yuan2015-da} shown in blue, PLCKG200.9-28.2 discussed by \citealp{Kale2017-if} shown in purple, A168 discussed by \citealp{Dwarakanath2018-ah} shown in orange, and A2018, A1904, A1697 discussed by \citealp{Van_Weeren2020-ua}  shown in black), and the masses from \citealp{Planck_Collaboration2016-jb}). The red rectangle marks the location of Abell~2108 with the mass uncertainties (X-ray vs. SZ, see Tab. \ref{tab:masses}) and extrapolation of the radio spectrum to 1.4\,GHz ($\alpha=-2$ and $\alpha=-1.2$).}
    \label{fig:power}
\end{figure*}
While we find reasonable agreement between the derived masses from SZ and the X-ray temperature for each component, the mass from the X-ray luminosity of Abell~2108 (east+west) is significantly lower than the SZ mass. 
The X-ray luminosity is more sensitive to the gas density than its temperature (see Tab. \ref{tab:masses}). Having reasonably good agreement between X-ray temperature and SZ derived masses suggests that the gas in Abell~2108 is heated, likely by the merger energy, and the X-ray luminosity does not reflect this. 
Since the mass from the X-ray luminosity of the background cluster agrees with its SZ mass, we conclude that Abell~2108 is X-ray underluminous due to shock heated gas.

If one had just the ROSAT/Planck catalog values at hand, one would clearly underestimate the mass of the field (Abell~2108 + Background cluster) in both cases. Attributing the entire SZ flux to one source at the same redshift as Abell~2108 (e.g., because the presence of the background cluster is not known), underestimates the mass by a factor of 2.4 ($>5\sigma$), while the mass from the X-ray flux underestimates the mass by a factor of 3.5 ($>12\sigma$), mainly due to the large distance of the background cluster primarily dimming the X-ray flux. 
SZ mass and X-ray luminosity-mass of the background cluster at the redshift of Abell~2108 (last row in Tab. \ref{tab:masses}) agree, indicating that it is not the presence of the background cluster that makes this an X-ray underluminous system, but the heating effects of the violent merger in Abell~2108. We conclude that our original approach, of selecting candidate cluster mergers based on X-ray underluminosity compared to their Planck SZ signal, is effective.

The mass estimates for merging clusters can easily be biased by the X-ray temperature or the SZ signal being affected through shock heating. 
Abell~1240 and Abell~3376 are two examples of clusters with known radio relics (\citealp{Botteon2020-co} and references therein) and a similar Planck SZ mass as Abell~2108. 
However, there are good indications that their true mass is actually higher: \cite{Barrena2009-dy} derive a dynamical mass for Abell~1240 of $\SIrange{1}{2e15}{M_\odot}$, and an ICM temperature measured with Chandra of around $\SI{6}{keV}$, much higher than expected for a $\SI{3.7e14}{M_\odot}$ system (the Planck SZ mass). 
Abell~3376 has Planck mass of $\SI{2.4e14}{M_\odot}$, while weak lensing measurements by \cite{Monteiro-Oliveira2017-ri} find a total mass of $\SI{4.7e14}{M_\odot}$. Also \cite{Durret2013-tb} confirmed a higher mass from galaxy velocity dispersion of $\SI{5.3e14}{M_\odot}$, and \cite{Urdampilleta2018-wx} find an ICM temperature with Suzaku of about $\SI{4.2}{keV}$. 
Weak lensing mass estimates are thought to be more reliable, and are frequently used to calibrate scaling relations. However, in the case of Abell~2108 the nearby background cluster will likely bias the weak lensing estimates, too. 
This may be solved through multi-component weak lensing analyses (\citealp{Okabe2014-wo}), but we note that also weak lensing mass estimates depend on the analysis strategy and include several systematic uncertainties often unaccounted (e.g., \citealp{Okabe2016-xq,Lovisari2020-ih}).

\subsection{Weak relics in low mass mergers}
With the wealth of newly available low-frequency radio data, more low mass systems with radio relics have been detected.
\cite{Dwarakanath2018-ah} found a relic in a low mass cluster well below $\SI{2e14}{M_\odot}$, and LOFAR detected several new candidates (because only at one frequency so far) in low mass merging clusters (\citealp{Van_Weeren2020-ua}). However, all of them are brighter in terms of the radio power at 1.4\,GHz than the relic in Abell~2108 (see Fig. \ref{fig:power}). 
For the comparison in Fig. \ref{fig:power} we had to extrapolate the radio flux of the relic in Abell~2108 from 383\,MHz to 1.4\,GHz using the measured spectral index of $-2$. If the higher frequency measurements (band 4 and even band 3) were missing significant flux due to poor uv coverage, a shallower spectral index would be measured and the radio power at 1.4\,GHz will be higher. However, even a spectral index of $-1.2$ (which is highly unlikely for Abell~2108) still yields a 1.4\,GHz radio power lower than any of the other objects shown in Fig. \ref{fig:power}. 
The relic radio power vs. cluster mass relation (Fig. \ref{fig:power} left, see also \citealp{De_Gasperin2014-qj}) has substantial scatter, but Abell~2108 seems to follow the trend, although at the lower end of radio power. The same is true for the radio power vs. largest linear size of the relic (Fig. \ref{fig:power} right). Clearly this newly detected radio feature in Abell~2108 - once all doubts about the higher frequency spectral behavior are clarified - marks the existence of weak radio relics from low Mach number shocks in low mass mergers with small Mach numbers. 
Although in the case of Abell~2108 the radio derived Mach number assuming DSA is close to the X-ray derived ones, the acceleration efficiency of such low Mach number shocks is expected to be too low to accelerate particles directly from the thermal plasma, and re-acceleration of seed relativistic electrons is more likely. Re-acceleration of relativistic particles still predicts agreement between the radio and X-ray Mach numbers (\citealp{Markevitch2005-ue}). 
Interestingly, this object was not found serendipitously, but through a combined X-ray/SZ selection of extreme clusters. The selection method will be demonstrated on more clusters in a forthcoming paper.  

\section{Summary}
\label{ch:summary}
Abell~2108 is a merger system consisting of two main subclusters, one at west being more luminous and regular, while the other at east has a highly disturbed morphology centered around a bright X-ray ridge.
The cluster mass based on the X-ray luminosity is $\SI{1.8e14}{M_\odot}$, while the SZ signal implies a 70\% higher mass. The east and west part of Abell~2108 have almost equal mass based on the X-ray temperature and SZ flux. 
The region between the cluster parts is faint in the X-rays, although they appear to be well past core passage. We do not detect the existence of a cool core in either subcluster. 
Although a nearby background cluster alters the signals stated in ROSAT X-ray and Planck SZ catalogs, that were used to derived cluster masses through scaling relations, we still find this cluster to be X-ray underluminous after taking the background cluster into account.

The western part of Abell~2108 shows several interesting features in the radio and X-rays: A temperature jump and surface brightness discontinuity, and extended radio emission consistent with a steep spectrum. 
This spectrum could be parameterized by a powerlaw, as expected if the emission is generated by DSA. However, the $\sim \SI{2}{\arcmin}$ spatial offset between the radio emission and the X-ray discontinuity implies that the shock is currently not (re-)accelerating relativistic electrons at the position of the radio feature.
Modeling the spectrum with a powerlaw and a spectral break yields a synchrotron cooling time estimate comparable to the inferred travel time of the shock from the radio emission to the X-ray discontinuity. 
This \textit{aged} radio relic is likely the result of the cluster merger, where a could of non-thermal particles has been re-accelerated in the past by the passage of a shock and is now fading, as suggested by its radio spectrum from 115 to 750\,MHz, and by the offset between the X-ray shock and the radio emission. 
Adiabatic compression of a non-thermal population of electrons could also explain a spectral steepening, but deeper radio data is needed to demonstrate this possibility. 
The exceptionally steep radio spectrum makes this relic the weakest relic known so far in terms of radio power. 

This finding demonstrates that low mass cluster mergers have the ability to host very steep spectrum radio relics, which are difficult to detect. 
Deeper X-ray observations will show the more clear features, such as the properties of the ICM between the east and west parts. 

\acknowledgments
{The authors thank the anonymous referee for the comments, which helped to improve the paper. 
GS acknowledges support through Chandra grant GO5-16126X. Basic research in radio astronomy at the Naval Research Laboratory is supported by 6.1 Base funding. 
SE and LL acknowledge financial contribution from the contracts ASI-INAF Athena 2019-27-HH.0, ``Attivit\`a di Studio per la comunit\`a scientifica di Astrofisica delle Alte Energie e Fisica Astroparticellare'' (Accordo Attuativo ASI-INAF n. 2017-14-H.0), INAF mainstream project 1.05.01.86.10, and
from the European Union’s Horizon 2020 Programme under the AHEAD2020 project (grant agreement n. 871158). 

DVL acknowledges the support of the Department of Atomic Energy, Government of India, under project no. 12-R\&D-TFR-5.02-0700.

We thank the staff of the GMRT who have made these observations possible. The GMRT is run by the National Centre for Radio Astrophysics of the Tata Institute of Fundamental Research. 

This publication makes is based on observations obtained with XMM-Newton, an ESA science mission with instruments and contributions directly funded by ESA Member States and NASA.
The scientific results reported in this article are based in part on observations made by the Chandra X-ray Observatory.}

\software{CIAO (v4.10, \citealp{Fruscione2006-wt}),
CASA (v5.6.0,  \citealp{McMullin2007-ed}), 
XSPEC (\citealp{Arnaud1996-uy}),
astropy (\citealp{The_Astropy_Collaboration2013-lw,The_Astropy_Collaboration2018-gx}), 
SAS (v15.0.0, \citealp{Gabriel2004-ty}),
Pyproffit (v0.5, \citealp{Eckert2020-xx})
PyBDSF (\citealp{Mohan2015-hz}),
SPAM (\citealp{Intema2009-cs})}

\clearpage
\bibliographystyle{aasjournal}
\bibliography{Paperpile.bib}

\begin{thebibliography}{}
\expandafter\ifx\csname natexlab\endcsname\relax\def\natexlab#1{#1}\fi
\providecommand{\url}[1]{\href{#1}{#1}}
\providecommand{\dodoi}[1]{doi:~\href{http://doi.org/#1}{\nolinkurl{#1}}}
\providecommand{\doeprint}[1]{\href{http://ascl.net/#1}{\nolinkurl{http://ascl.net/#1}}}
\providecommand{\doarXiv}[1]{\href{https://arxiv.org/abs/#1}{\nolinkurl{https://arxiv.org/abs/#1}}}

\bibitem[{Abell(1958)}]{Abell1958-dx}
Abell, G.~O. 1958, \apjs, 3, 211

\bibitem[{Alam {et~al.}(2015)Alam, Albareti, Prieto, Anders, Anderson,
  Anderton, Andrews, Armengaud, Aubourg, Bailey, Basu, Bautista, Beaton, Beers,
  Bender, Berlind, Beutler, Bhardwaj, Bird, Bizyaev, Blake, Blanton, Blomqvist,
  Bochanski, Bolton, Bovy, Shelden~Bradley, Brandt, Brauer, Brinkmann, Brown,
  Brownstein, Burden, Burtin, Busca, Cai, Capozzi, Rosell, Carr, Carrera,
  Chambers, Chaplin, Chen, Chiappini, Drew~Chojnowski, Chuang, Clerc, Comparat,
  Covey, Croft, Cuesta, Cunha, da~Costa, Da~Rio, Davenport, Dawson, De~Lee,
  Delubac, Deshpande, Dhital, Dutra-Ferreira, Dwelly, Ealet, Ebelke, Edmondson,
  Eisenstein, Ellsworth, Elsworth, Epstein, Eracleous, Escoffier, Esposito,
  Evans, Fan, Fern{\'a}ndez-Alvar, Feuillet, Ak, Finley, Finoguenov, Flaherty,
  Fleming, Font-Ribera, Foster, Frinchaboy, Galbraith-Frew, Garc{\'\i}a,
  Garc{\'\i}a-Hern{\'a}ndez, Garc{\'\i}a~P{\'e}rez, Gaulme, Ge,
  G{\'e}nova-Santos, Georgakakis, Ghezzi, Gillespie, Girardi, Goddard, Gontcho,
  Gonz{\'a}lez~Hern{\'a}ndez, Grebel, Green, Grieb, Grieves, Gunn, Guo,
  Harding, Hasselquist, Hawley, Hayden, Hearty, Hekker, Ho, Hogg,
  Holley-Bockelmann, Holtzman, Honscheid, Huber, Huehnerhoff, Ivans, Jiang,
  Johnson, Kinemuchi, Kirkby, Kitaura, Klaene, Knapp, Kneib, Koenig, Lam, Lan,
  Lang, Laurent, Le~Goff, Leauthaud, Lee, Lee, Licquia, Liu, Long,
  L{\'o}pez-Corredoira, Lorenzo-Oliveira, Lucatello, Lundgren, Lupton, Mack,
  Mahadevan, Maia, Majewski, Malanushenko, Malanushenko, Manchado, Manera, Mao,
  Maraston, Marchwinski, Margala, Martell, Martig, Masters, Mathur, McBride,
  McGehee, McGreer, McMahon, M{\'e}nard, Menzel, Merloni, M{\'e}sz{\'a}ros,
  Miller, Miralda-Escud{\'e}, Miyatake, Montero-Dorta, More, Morganson,
  Morice-Atkinson, Morrison, Mosser, Muna, Myers, Nandra, Newman, Neyrinck,
  Nguyen, Nichol, Nidever, Noterdaeme, Nuza, O'Connell, O'Connell, O'Connell,
  Ogando, Olmstead, Oravetz, Oravetz, Osumi, Owen, Padgett, Padmanabhan,
  Paegert, Palanque-Delabrouille, Pan, Parejko, P{\^a}ris, Park,
  Pattarakijwanich, Pellejero-Ibanez, Pepper, Percival, P{\'e}rez-Fournon,
  Pe´rez-Ra`fols, Petitjean, Pieri, Pinsonneault, de~Mello, Prada, Prakash,
  Price-Whelan, Protopapas, Jordan~Raddick, Rahman, Reid, Rich, Rix, Robin,
  Rockosi, Rodrigues, Rodr{\'\i}guez-Torres, Roe, Ross, Ross, Rossi, Ruan,
  Rubi{\~n}o-Mart{\'\i}n, Rykoff, Salazar-Albornoz, Salvato, Samushia,
  S{\'a}nchez, Santiago, Sayres, Schiavon, Schlegel, Schmidt, Schneider,
  Schultheis, Schwope, Sc{\'o}ccola, Scott, Sellgren, Seo, Serenelli, Shane,
  Shen, Shetrone, Shu, Silva~Aguirre, Sivarani, Skrutskie, Slosar, Smith,
  Sobreira, Souto, Stassun, Steinmetz, Stello, Strauss, Streblyanska, Suzuki,
  Swanson, Tan, Tayar, Terrien, Thakar, Thomas, Thomas, Thompson, Tinker,
  Tojeiro, Troup, Vargas-Maga{\~n}a, Vazquez, Verde, Viel, Vogt, Wake, Wang,
  Weaver, Weinberg, Weiner, White, Wilson, Wisniewski, Wood-Vasey, Ye`che,
  York, Zakamska, Zamora, Zasowski, Zehavi, Zhao, Zheng, (周旭), (周志民),
  (邹虎), \& Zhu}]{Alam2015-yc}
Alam, S., Albareti, F.~D., Prieto, C.~A., {et~al.} 2015, ApJS, 219, 12

\bibitem[{Arnaud(1996)}]{Arnaud1996-uy}
Arnaud, K.~A. 1996, in Astronomical Society of the Pacific Conference Series,
  Vol. 101, Astronomical Data Analysis Software and Systems {V}, ed. {G. H.
  Jacoby \& J. Barnes}, 17

\bibitem[{Arnaud {et~al.}(2010)Arnaud, Pratt, Piffaretti, B{\"o}hringer,
  Croston, \& Pointecouteau}]{Arnaud2010-ct}
Arnaud, M., Pratt, G.~W., Piffaretti, R., {et~al.} 2010, Astron. Astrophys.
  Suppl. Ser., 517, A92

\bibitem[{Arnaud {et~al.}(2021)Arnaud, Ettori, Pratt, Rossetti, Eckert,
  Gastaldello, Gavazzi, Kay, Lovisari, Maughan, Pointecouteau, Sereno,
  Bartalucci, Bonafede, Bourdin, Cassano, Duffy, Iqbal, Maurogordato, Rasia,
  Sayers, Andrade-Santos, Aussel, Barnes, Barrena, Borgani, Burkutean, Clerc,
  Corasaniti, Cuillandre, De~Grandi, De~Petris, Dolag, Donahue, Ferragamo,
  Gaspari, Ghizzardi, Gitti, Haines, Jauzac, Johnston-Hollitt, Jones,
  K{\'e}ruzor{\'e}, LeBrun, Mayet, Mazzotta, Melin, Molendi, Nonino, Okabe,
  Paltani, Perotto, Pires, Radovich, Rubino-Martin, Salvati, Saro, Sartoris,
  Schellenberger, Streblyanska, Tarr{\'\i}o, Tozzi, Umetsu, van~der Burg,
  Vazza, Venturi, Yepes, \& Zarattini}]{Arnaud2021-nz}
Arnaud, M., Ettori, S., Pratt, G.~W., {et~al.} 2021, Astron. Astrophys. Suppl.
  Ser., 650, A104

\bibitem[{Barrena {et~al.}(2009)Barrena, Girardi, Boschin, \&
  Das{\'\i}}]{Barrena2009-dy}
Barrena, R., Girardi, M., Boschin, W., \& Das{\'\i}, M. 2009, Astron.
  Astrophys. Suppl. Ser., 503, 357

\bibitem[{Botteon {et~al.}(2020)Botteon, Brunetti, Ryu, \&
  Roh}]{Botteon2020-co}
Botteon, A., Brunetti, G., Ryu, D., \& Roh, S. 2020, Astron. Astrophys. Suppl.
  Ser., 634, A64

\bibitem[{Brunetti \& Jones(2014)}]{Brunetti2014-lw}
Brunetti, G., \& Jones, T.~W. 2014, Int. J. Mod. Phys. D, 23, 1430007

\bibitem[{De~Gasperin {et~al.}(2014)De~Gasperin, van Weeren, Br{\"u}ggen,
  Vazza, Bonafede, \& Intema}]{De_Gasperin2014-qj}
De~Gasperin, F., van Weeren, R.~J., Br{\"u}ggen, M., {et~al.} 2014, Mon. Not.
  R. Astron. Soc., 444, 3130

\bibitem[{Donnert {et~al.}(2018)Donnert, Vazza, Br{\"u}ggen, \&
  ZuHone}]{Donnert2018-an}
Donnert, J., Vazza, F., Br{\"u}ggen, M., \& ZuHone, J. 2018, Space Sci. Rev.,
  214, 122

\bibitem[{Durret {et~al.}(2013)Durret, Perrot, Lima~Neto, Adami, Bertin, \&
  Bagchi}]{Durret2013-tb}
Durret, F., Perrot, C., Lima~Neto, G.~B., {et~al.} 2013, Astron. Astrophys.
  Suppl. Ser., 560, A78

\bibitem[{Dwarakanath {et~al.}(2018)Dwarakanath, Parekh, Kale, \&
  George}]{Dwarakanath2018-ah}
Dwarakanath, K.~S., Parekh, V., Kale, R., \& George, L.~T. 2018, Mon. Not. R.
  Astron. Soc., 477, 957

\bibitem[{Ebeling {et~al.}(1998)Ebeling, Edge, B{\"o}hringer, Allen, Crawford,
  Fabian, Voges, \& Huchra}]{Ebeling1998-za}
Ebeling, H., Edge, A.~C., B{\"o}hringer, H., {et~al.} 1998, Mon. Not. R.
  Astron. Soc., 301, 881

\bibitem[{Eckert {et~al.}(2020)Eckert, Finoguenov, Ghirardini, Grandis, Kaefer,
  Sanders, \& Ramos-Ceja}]{Eckert2020-xx}
Eckert, D., Finoguenov, A., Ghirardini, V., {et~al.} 2020.
\newblock \doarXiv{2009.03944}

\bibitem[{Eckert {et~al.}(2021)Eckert, Gaspari, Gastaldello, Le~Brun, \&
  O'Sullivan}]{Eckert2021-mm}
Eckert, D., Gaspari, M., Gastaldello, F., Le~Brun, A. M.~C., \& O'Sullivan, E.
  2021, Universe, 7, 142

\bibitem[{En{\ss}lin \& {Gopal-Krishna}(2001)}]{Enslin2001-jg}
En{\ss}lin, T.~A., \& {Gopal-Krishna}. 2001, Astron. Astrophys. Suppl. Ser.,
  366, 26

\bibitem[{Forbes {et~al.}(2006)Forbes, Ponman, Pearce, Osmond, Kilborn, Brough,
  Raychaudhury, Mundell, Miles, \& Kern}]{Forbes2006-ay}
Forbes, D.~A., Ponman, T., Pearce, F., {et~al.} 2006, Publications of the
  Astronomical Society of Australia, 23, 38

\bibitem[{Fruscione {et~al.}(2006)Fruscione, McDowell, Allen, \&
  {others}}]{Fruscione2006-wt}
Fruscione, A., McDowell, J.~C., Allen, G.~E., \& {others}. 2006, Observatory

\bibitem[{Gabriel {et~al.}(2004)Gabriel, Denby, Fyfe, Hoar, \&
  {others}}]{Gabriel2004-ty}
Gabriel, C., Denby, M., Fyfe, D.~J., Hoar, J., \& {others}. 2004, Astronomical
  Data

\bibitem[{Gastaldello {et~al.}(2013)Gastaldello, Di~Gesu, Ghizzardi,
  Giacintucci, Girardi, Roediger, Rossetti, Brighenti, Buote, Eckert, Ettori,
  Humphrey, \& Mathews}]{Gastaldello2013-cp}
Gastaldello, F., Di~Gesu, L., Ghizzardi, S., {et~al.} 2013, ApJ, 770, 56

\bibitem[{Giacintucci {et~al.}(2008)Giacintucci, Venturi, Macario, Dallacasa,
  Brunetti, Markevitch, Cassano, Bardelli, \& Athreya}]{Giacintucci2008-dx}
Giacintucci, S., Venturi, T., Macario, G., {et~al.} 2008, Astron. Astrophys.
  Suppl. Ser., 486, 347

\bibitem[{Golovich {et~al.}(2019)Golovich, Dawson, Wittman, van Weeren,
  Andrade-Santos, Jee, Benson, de~Gasperin, Venturi, Bonafede, Sobral, Ogrean,
  Lemaux, Brada{\v c}, Br{\"u}ggen, \& Peter}]{Golovich2019-zd}
Golovich, N., Dawson, W.~A., Wittman, D.~M., {et~al.} 2019, ApJ, 882, 69

\bibitem[{Govoni \& Feretti(2004)}]{Govoni2004-dx}
Govoni, F., \& Feretti, L. 2004, Int. J. Mod. Phys. D, 13, 1549

\bibitem[{Ha {et~al.}(2018)Ha, Ryu, \& Kang}]{Ha2018-ou}
Ha, J.-H., Ryu, D., \& Kang, H. 2018, ApJ, 857, 26

\bibitem[{Hoang {et~al.}(2017)Hoang, Shimwell, Stroe, Akamatsu, Brunetti,
  Donnert, Intema, Mulcahy, R{\"o}ttgering, van Weeren, \&
  {Others}}]{Hoang2017-pu}
Hoang, D.~N., Shimwell, T.~W., Stroe, A., {et~al.} 2017, Mon. Not. R. Astron.
  Soc., 471, 1107

\bibitem[{Hoeft \& Br{\"u}ggen(2007)}]{Hoeft2007-cv}
Hoeft, M., \& Br{\"u}ggen, M. 2007, Mon. Not. R. Astron. Soc., 375, 77

\bibitem[{Intema {et~al.}(2009)Intema, van~der Tol, Cotton, Cohen, van Bemmel,
  \& R{\"o}ttgering}]{Intema2009-cs}
Intema, H.~T., van~der Tol, S., Cotton, W.~D., {et~al.} 2009, \aap, 501, 1185

\bibitem[{Kale \& Dwarakanath(2009)}]{Kale2009-yx}
Kale, R., \& Dwarakanath, K.~S. 2009, ApJ, 699, 1883

\bibitem[{Kale \& Dwarakanath(2012)}]{Kale2012-ok}
---. 2012, Astrophys. J., 744, 46

\bibitem[{Kale {et~al.}(2017)Kale, Wik, Giacintucci, Venturi, Brunetti,
  Cassano, Dallacasa, \& de~Gasperin}]{Kale2017-if}
Kale, R., Wik, D.~R., Giacintucci, S., {et~al.} 2017, Mon. Not. R. Astron.
  Soc., 472, 940

\bibitem[{Kang \& Ryu(2013)}]{Kang2013-nf}
Kang, H., \& Ryu, D. 2013, ApJ, 764, 95

\bibitem[{Kempner \& Sarazin(2001)}]{Kempner2001-eb}
Kempner, J.~C., \& Sarazin, C.~L. 2001, ApJ, 548, 639

\bibitem[{Kraft {et~al.}(2006)Kraft, Jones, Nulsen, \&
  Hardcastle}]{Kraft2006-wn}
Kraft, R.~P., Jones, C., Nulsen, P. E.~J., \& Hardcastle, M.~J. 2006,
  Astrophys. J., 640, 762

\bibitem[{Landau \& Lifshitz(1959)}]{Landau1959-li}
Landau, L.~D., \& Lifshitz, E.~M. 1959, Course of Theoretical Physics Vol. 6
  Fluid Mechanies (Pergamon Press)

\bibitem[{Lin {et~al.}(2018)Lin, Huang, \& Chen}]{Lin2018-jt}
Lin, Y.-T., Huang, H.-J., \& Chen, Y.-C. 2018, AJS, 155, 188

\bibitem[{Locatelli {et~al.}(2020)Locatelli, Rajpurohit, Vazza, Gastaldello,
  Dallacasa, Bonafede, Rossetti, Stuardi, Bonassieux, Brunetti, \&
  {Others}}]{Locatelli2020-cs}
Locatelli, N.~T., Rajpurohit, K., Vazza, F., {et~al.} 2020, Mon. Not. R. Aston.
  Soc. Lett., 496, L48

\bibitem[{Lovisari {et~al.}(2021)Lovisari, Ettori, Gaspari, \&
  Giles}]{Lovisari2021-ix}
Lovisari, L., Ettori, S., Gaspari, M., \& Giles, P.~A. 2021, Universe, 7, 139

\bibitem[{Lovisari {et~al.}(2020{\natexlab{a}})Lovisari, Ettori, Sereno,
  Schellenberger, Forman, Andrade-Santos, \& Jones}]{Lovisari2020-ih}
Lovisari, L., Ettori, S., Sereno, M., {et~al.} 2020{\natexlab{a}}, Astron.
  Astrophys. Suppl. Ser., 644, A78

\bibitem[{Lovisari \& Reiprich(2019)}]{Lovisari2019-nh}
Lovisari, L., \& Reiprich, T.~H. 2019, Mon. Not. R. Astron. Soc., 483, 540

\bibitem[{Lovisari {et~al.}(2015)Lovisari, Reiprich, \&
  Schellenberger}]{Lovisari2015-xm}
Lovisari, L., Reiprich, T.~H., \& Schellenberger, G. 2015, Astron. Astrophys.
  Suppl. Ser., 573, A118

\bibitem[{Lovisari {et~al.}(2020{\natexlab{b}})Lovisari, Schellenberger,
  Sereno, Ettori, Pratt, Forman, Jones, Andrade-Santos, Randall, \&
  Kraft}]{Lovisari2020-gn}
Lovisari, L., Schellenberger, G., Sereno, M., {et~al.} 2020{\natexlab{b}}, ApJ,
  892, 102

\bibitem[{Machacek {et~al.}(2011)Machacek, Jerius, Kraft, Forman, Jones,
  Randall, Giacintucci, \& Sun}]{Machacek2011-po}
Machacek, M.~E., Jerius, D., Kraft, R., {et~al.} 2011, ApJ, 743, 15

\bibitem[{Machacek {et~al.}(2005)Machacek, Nulsen, Stirbat, Jones, \&
  Forman}]{Machacek2005-el}
Machacek, M.~E., Nulsen, P., Stirbat, L., Jones, C., \& Forman, W.~R. 2005,
  \apj, 630, 280

\bibitem[{Machacek {et~al.}(2010)Machacek, O'Sullivan, Randall, Jones, \&
  Forman}]{Machacek2010-qu}
Machacek, M.~E., O'Sullivan, E., Randall, S.~W., Jones, C., \& Forman, W.~R.
  2010, Astrophys. J., 711, 1316

\bibitem[{Machado \& Lima~Neto(2013)}]{Machado2013-em}
Machado, R. E.~G., \& Lima~Neto, G.~B. 2013, Mon. Not. R. Astron. Soc., 430,
  3249

\bibitem[{Markevitch {et~al.}(2005)Markevitch, Govoni, Brunetti, \&
  Jerius}]{Markevitch2005-ue}
Markevitch, M., Govoni, F., Brunetti, G., \& Jerius, D. 2005, ApJ, 627, 733

\bibitem[{McMullin {et~al.}(2007)McMullin, Waters, Schiebel, Young, \&
  Golap}]{McMullin2007-ed}
McMullin, J.~P., Waters, B., Schiebel, D., Young, W., \& Golap, K. 2007, in
  Astronomical Society of the Pacific Conference Series, Vol. 376, Astronomical
  Data Analysis Software and Systems {XVI}, ed. R.~A. Shaw, F.~Hill, \& D.~J.
  Bell, 127

\bibitem[{Merloni {et~al.}(2012)Merloni, Predehl, Becker, B{\"o}hringer,
  Boller, Brunner, Brusa, Dennerl, Freyberg, Friedrich, Georgakakis, Haberl,
  Hasinger, Meidinger, Mohr, Nandra, Rau, Reiprich, Robrade, Salvato,
  Santangelo, Sasaki, Schwope, Wilms, \& {the German eROSITA
  Consortium}}]{Merloni2012-sc}
Merloni, A., Predehl, P., Becker, W., {et~al.} 2012.
\newblock \doarXiv{1209.3114}

\bibitem[{Mohan \& Rafferty(2015)}]{Mohan2015-hz}
Mohan, N., \& Rafferty, D. 2015, {PyBDSF}: Python Blob Detection and Source
  Finder, Astrophysics Source Code Library

\bibitem[{Monteiro-Oliveira {et~al.}(2017)Monteiro-Oliveira, Lima~Neto,
  Cypriano, Machado, Capelato, Lagan{\'a}, Durret, \&
  Bagchi}]{Monteiro-Oliveira2017-ri}
Monteiro-Oliveira, R., Lima~Neto, G.~B., Cypriano, E.~S., {et~al.} 2017, Mon.
  Not. R. Astron. Soc., 468, 4566

\bibitem[{Nuza {et~al.}(2017)Nuza, Gelszinnis, Hoeft, \& Yepes}]{Nuza2017-ir}
Nuza, S.~E., Gelszinnis, J., Hoeft, M., \& Yepes, G. 2017, Mon. Not. R. Astron.
  Soc., 470, 240

\bibitem[{Offringa {et~al.}(2014)Offringa, McKinley, Hurley-Walker, Briggs,
  Wayth, Kaplan, Bell, Feng, Neben, Hughes, Rhee, Murphy, Bhat, Bernardi,
  Bowman, Cappallo, Corey, Deshpande, Emrich, Ewall-Wice, Gaensler, Goeke,
  Greenhill, Hazelton, Hindson, Johnston-Hollitt, Jacobs, Kasper, Kratzenberg,
  Lenc, Lonsdale, Lynch, McWhirter, Mitchell, Morales, Morgan, Kudryavtseva,
  Oberoi, Ord, Pindor, Procopio, Prabu, Riding, Roshi, Shankar, Srivani,
  Subrahmanyan, Tingay, Waterson, Webster, Whitney, Williams, \&
  Williams}]{Offringa2014-bw}
Offringa, A.~R., McKinley, B., Hurley-Walker, N., {et~al.} 2014, Mon. Not. R.
  Astron. Soc., 444, 606

\bibitem[{Ogrean {et~al.}(2013)Ogrean, Br{\"u}ggen, van Weeren, R{\"o}ttgering,
  Croston, \& Hoeft}]{Ogrean2013-bi}
Ogrean, G.~A., Br{\"u}ggen, M., van Weeren, R.~J., {et~al.} 2013, \mnras, 433,
  812

\bibitem[{Okabe {et~al.}(2014)Okabe, Futamase, Kajisawa, \&
  Kuroshima}]{Okabe2014-wo}
Okabe, N., Futamase, T., Kajisawa, M., \& Kuroshima, R. 2014, Astrophys. J.,
  784, 90

\bibitem[{Okabe \& Smith(2016)}]{Okabe2016-xq}
Okabe, N., \& Smith, G.~P. 2016, \mnras, 461, 3794

\bibitem[{O'Sullivan {et~al.}(2019)O'Sullivan, Schellenberger, Burke, Sun,
  Vrtilek, David, \& Sarazin}]{OSullivan2019-le}
O'Sullivan, E., Schellenberger, G., Burke, D.~J., {et~al.} 2019, Mon. Not. R.
  Astron. Soc., 488, 2925

\bibitem[{O'Sullivan {et~al.}(2014)O'Sullivan, Vrtilek, David, Giacintucci,
  Zezas, Ponman, Mamon, Nulsen, \& Raychaudhury}]{OSullivan2014-ku}
O'Sullivan, E., Vrtilek, J.~M., David, L.~P., {et~al.} 2014, \apj, 793, 74

\bibitem[{Piffaretti {et~al.}(2011)Piffaretti, Arnaud, Pratt, Pointecouteau, \&
  Melin}]{Piffaretti2011-ay}
Piffaretti, R., Arnaud, M., Pratt, G.~W., Pointecouteau, E., \& Melin, J.-B.
  2011, {\aa}, 534, A109

\bibitem[{{Planck Collaboration} {et~al.}(2016){Planck Collaboration}, {Ade, P
  A R}, Aghanim, Arnaud, Ashdown, \& {others}}]{Planck_Collaboration2016-jb}
{Planck Collaboration}, {Ade, P A R}, Aghanim, N., {et~al.} 2016, Astronomy

\bibitem[{Popesso {et~al.}(2007)Popesso, Biviano, B{\"o}hringer, \&
  Romaniello}]{Popesso2007-hp}
Popesso, P., Biviano, A., B{\"o}hringer, H., \& Romaniello, M. 2007, Astron.
  Astrophys. Suppl. Ser., 461, 397

\bibitem[{Randall {et~al.}(2009)Randall, Jones, Kraft, Forman, \&
  O'Sullivan}]{Randall2009-dn}
Randall, S.~W., Jones, C., Kraft, R., Forman, W.~R., \& O'Sullivan, E. 2009,
  Astrophys. J., 696, 1431

\bibitem[{Richards {et~al.}(2015)Richards, Myers, Peters, Krawczyk, Chase,
  Ross, Fan, Jiang, Lacy, McGreer, Trump, \& Riegel}]{Richards2015-ae}
Richards, G.~T., Myers, A.~D., Peters, C.~M., {et~al.} 2015, ApJS, 219, 39

\bibitem[{Russell {et~al.}(2014)Russell, Fabian, McNamara, Edge, Sanders,
  Nulsen, Baum, Donahue, \& O'Dea}]{Russell2014-zj}
Russell, H.~R., Fabian, A.~C., McNamara, B.~R., {et~al.} 2014, Mon. Not. R.
  Astron. Soc., 444, 629

\bibitem[{Sarazin {et~al.}(1982)Sarazin, Rood, \& Struble}]{Sarazin1982-az}
Sarazin, C.~L., Rood, H.~J., \& Struble, M.~F. 1982, Astron. Astrophys.

\bibitem[{Schellenberger \& Reiprich(2017)}]{Schellenberger2017-mc}
Schellenberger, G., \& Reiprich, T.~H. 2017, Mon. Not. R. Astron. Soc., 469,
  3738

\bibitem[{Shimizu {et~al.}(2003)Shimizu, Kitayama, Sasaki, \&
  Suto}]{Shimizu2003-nv}
Shimizu, M., Kitayama, T., Sasaki, S., \& Suto, Y. 2003, ApJ, 590, 197

\bibitem[{Slee {et~al.}(2001)Slee, Roy, Murgia, Andernach, \&
  Ehle}]{Slee2001-jw}
Slee, O.~B., Roy, A.~L., Murgia, M., Andernach, H., \& Ehle, M. 2001, AJS, 122,
  1172

\bibitem[{Sun(2009)}]{Sun2009-pg}
Sun, M. 2009, \apj, 704, 1586

\bibitem[{Sun(2012)}]{Sun2012-ll}
---. 2012, New J. Phys., 14, 045004

\bibitem[{{The Astropy Collaboration} {et~al.}(2013){The Astropy
  Collaboration}, Robitaille, Tollerud, Greenfield, Droettboom, Bray, Aldcroft,
  Davis, Ginsburg, Price-Whelan, Kerzendorf, Conley, Crighton, Barbary, Muna,
  Ferguson, Grollier, Parikh, Nair, G{\"u}nther, Deil, Woillez, Conseil,
  Kramer, Turner, Singer, Fox, Weaver, Zabalza, Edwards, Azalee~Bostroem,
  Burke, Casey, Crawford, Dencheva, Ely, Jenness, Labrie, Lim, Pierfederici,
  Pontzen, Ptak, Refsdal, Servillat, \&
  Streicher}]{The_Astropy_Collaboration2013-lw}
{The Astropy Collaboration}, Robitaille, T.~P., Tollerud, E.~J., {et~al.} 2013,
  Astron. Astrophys. Suppl. Ser., 558, A33

\bibitem[{{The Astropy Collaboration} {et~al.}(2018){The Astropy
  Collaboration}, Price-Whelan, Sip{\H o}cz, G{\"u}nther, Lim, Crawford,
  Conseil, Shupe, Craig, Dencheva, Ginsburg, VanderPlas, Bradley,
  P{\'e}rez-Su{\'a}rez, de~Val-Borro, Aldcroft, Cruz, Robitaille, Tollerud,
  Ardelean, Babej, Bach, Bachetti, Bakanov, Bamford, Barentsen, Barmby,
  Baumbach, Berry, Biscani, Boquien, Bostroem, Bouma, Brammer, Bray,
  Breytenbach, Buddelmeijer, Burke, Calderone, Cano~Rodr{\'\i}guez, Cara,
  Cardoso, Cheedella, Copin, Corrales, Crichton, D'Avella, Deil, Depagne,
  Dietrich, Donath, Droettboom, Earl, Erben, Fabbro, Ferreira, Finethy, Fox,
  Garrison, Gibbons, Goldstein, Gommers, Greco, Greenfield, Groener, Grollier,
  Hagen, Hirst, Homeier, Horton, Hosseinzadeh, Hu, Hunkeler, Ivezi{\'c}, Jain,
  Jenness, Kanarek, Kendrew, Kern, Kerzendorf, Khvalko, King, Kirkby, Kulkarni,
  Kumar, Lee, Lenz, Littlefair, Ma, Macleod, Mastropietro, McCully, Montagnac,
  Morris, Mueller, Mumford, Muna, Murphy, Nelson, Nguyen, Ninan, N{\"o}the,
  Ogaz, Oh, Parejko, Parley, Pascual, Patil, Patil, Plunkett, Prochaska,
  Rastogi, Reddy~Janga, Sabater, Sakurikar, Seifert, Sherbert, Sherwood-Taylor,
  Shih, Sick, Silbiger, Singanamalla, Singer, Sladen, Sooley, Sornarajah,
  Streicher, Teuben, Thomas, Tremblay, Turner, Terr{\'o}n, van Kerkwijk, de~la
  Vega, Watkins, Weaver, Whitmore, Woillez, Zabalza, {(Primary Paper
  Contributors)}, {(Astropy Coordination Committee)}, \& {(Astropy
  Contributors)}}]{The_Astropy_Collaboration2018-gx}
{The Astropy Collaboration}, Price-Whelan, A.~M., Sip{\H o}cz, B.~M., {et~al.}
  2018, Astron. J., 156, 123

\bibitem[{Urdampilleta {et~al.}(2018)Urdampilleta, Akamatsu, Mernier, Kaastra,
  de~Plaa, Ohashi, Ishisaki, \& Kawahara}]{Urdampilleta2018-wx}
Urdampilleta, I., Akamatsu, H., Mernier, F., {et~al.} 2018, Astron. Astrophys.
  Suppl. Ser., 618, A74

\bibitem[{van Weeren {et~al.}(2019)van Weeren, de~Gasperin, Akamatsu,
  Br{\"u}ggen, Feretti, Kang, Stroe, \& Zandanel}]{Van_Weeren2019-qg}
van Weeren, R.~J., de~Gasperin, F., Akamatsu, H., {et~al.} 2019, Space Sci.
  Rev., 215, 16

\bibitem[{Van~Weeren {et~al.}(2009)Van~Weeren, R{\"o}ttgering, Br{\"u}ggen, \&
  Cohen}]{Van_Weeren2009-sn}
Van~Weeren, R.~J., R{\"o}ttgering, H. J.~A., Br{\"u}ggen, M., \& Cohen, A.
  2009, Astron. Astrophys. Suppl. Ser., 508, 75

\bibitem[{van Weeren {et~al.}(2016)van Weeren, Brunetti, Br{\"u}ggen,
  Andrade-Santos, Ogrean, Williams, R{\"o}ttgering, Dawson, Forman, Gasperin,
  Hardcastle, Jones, Miley, Rafferty, Rudnick, Sabater, Sarazin, Shimwell,
  Bonafede, Best, Bîrzan, Cassano, Chy{\.z}y, Croston, Dijkema, En{\ss}lin,
  Ferrari, Heald, Hoeft, Horellou, Jarvis, Kraft, Mevius, Intema, Murray,
  Orr{\'u}, Pizzo, Sridhar, Simionescu, Stroe, Tol, \&
  White}]{Van_Weeren2016-gt}
van Weeren, R.~J., Brunetti, G., Br{\"u}ggen, M., {et~al.} 2016, Astrophys. J.,
  818, 204

\bibitem[{van Weeren {et~al.}(2017)van Weeren, Andrade-Santos, Dawson,
  Golovich, Lal, Kang, Ryu, Br{\"u}ggen, Ogrean, Forman, Jones, Placco,
  Santucci, Wittman, Jee, Kraft, Sobral, Stroe, \& Fogarty}]{Van_Weeren2017-ej}
van Weeren, R.~J., Andrade-Santos, F., Dawson, W.~A., {et~al.} 2017, Nature
  Astronomy, 1, 0005

\bibitem[{van Weeren {et~al.}(2020)van Weeren, Shimwell, Botteon, Brunetti,
  Br{\"u}ggen, Boxelaar, Cassano, Di~Gennaro, Andrade-Santos, Bonnassieux,
  Bonafede, Cuciti, Dallacasa, de~Gasperin, Gastaldello, Hardcastle, Hoeft,
  Kraft, Mandal, Rossetti, R{\"o}ttgering, Tasse, \&
  Wilber}]{Van_Weeren2020-ua}
van Weeren, R.~J., Shimwell, T.~W., Botteon, A., {et~al.} 2020.
\newblock \doarXiv{2011.02387}

\bibitem[{Vazza \& Br{\"u}ggen(2014)}]{Vazza2014-ta}
Vazza, F., \& Br{\"u}ggen, M. 2014, Mon. Not. R. Astron. Soc., 437, 2291

\bibitem[{Wen \& Han(2013)}]{Wen2013-rf}
Wen, Z.~L., \& Han, J.~L. 2013, Mon. Not. R. Astron. Soc., 436, 275

\bibitem[{Wen {et~al.}(2010)Wen, Han, \& Liu}]{Wen2010-hl}
Wen, Z.~L., Han, J.~L., \& Liu, F.~S. 2010, Astrophys. J. Suppl. Ser., 187, 272

\bibitem[{Yuan {et~al.}(2015)Yuan, Han, \& Wen}]{Yuan2015-da}
Yuan, Z.~S., Han, J.~L., \& Wen, Z.~L. 2015, Astrophys. J., 813, 77

\bibitem[{ZuHone {et~al.}(2021)ZuHone, Markevitch, Weinberger, Nulsen, \&
  Ehlert}]{ZuHone2020-rp}
ZuHone, J.~A., Markevitch, M., Weinberger, R., Nulsen, P., \& Ehlert, K. 2021,
  ApJ, 914, 73

\end{thebibliography}

\end{document}